\documentclass[manuscript,screen,nonacm]{acmart}

\AtBeginDocument{%
  \providecommand\BibTeX{{%
    \normalfont B\kern-0.5em{\scshape i\kern-0.25em b}\kern-0.8em\TeX}}}

\renewcommand\footnotetextcopyrightpermission[1]{} 
\setcopyright{none}
\usepackage[normalem]{ulem}

\usepackage{float}
\usepackage{booktabs} 
\usepackage{multirow}

\usepackage{graphicx}

\usepackage{tabularx}
\newcolumntype{Y}{>{\centering\arraybackslash}X}
\usepackage{diagbox}
\usepackage{makecell}

\newcommand{\framework}{AutoDSE}
\newcommand{\speedupcpu}{19.9}

\newcommand{\pragmareductionvitis}{26.38$\times$}

\usepackage{tikz}
\newcommand*\circled[1]{\tikz[baseline=(char.base)]{
            \node[shape=circle,draw,inner sep=0.5pt] (char) {#1};}}
\makeatletter
\global\let\tikz@ensure@dollar@catcode=\relax
\makeatother            

\usepackage{layouts}

\usepackage{caption}
\usepackage{subcaption}

\usepackage{amsmath}
\newcommand{\myquote}[1]{\ensuremath{\text{\textquotedbl} #1 \text{\textquotedbl}}}
\usepackage{etoolbox}

\usepackage{titlesec}
\usepackage{lipsum}

\usepackage[colorinlistoftodos]{todonotes}
\usepackage{algorithm}
\usepackage{algorithmic}
\usepackage{xcolor}

\usepackage{balance}
\usepackage{listings}
\usepackage{color}
\definecolor{dkgreen}{rgb}{0,0.6,0}
\definecolor{gray}{rgb}{0.5,0.5,0.5}
\definecolor{mauve}{rgb}{0.58,0,0.82}
\definecolor{bg}{rgb}{0.9,0.9,0.9}

\begin{document}

\title{
AutoDSE: Enabling Software Programmers to Design Efficient FPGA Accelerators}

\author{Atefeh Sohrabizadeh}
\authornote{Both authors contributed equally to this research.}
\email{atefehsz@cs.ucla.edu}
\affiliation{%
  \institution{Computer Science Department, University of California, Los Angeles}
  \city{Los Angeles}
  \state{CA}
  \country{USA}
}

\author{Cody Hao Yu}
\authornotemark[1]
\email{hyu@cs.ucla.edu}
\affiliation{%
  \institution{Computer Science Department, University of California, Los Angeles}
  \city{Los Angeles}
  \state{CA}
  \country{USA}
}

\author{Min Gao}
\affiliation{%
  \institution{Falcon-computing Inc.}
  \city{Los Angeles}
  \country{USA}}
\email{mingao@falcon-computing.com}

\author{Jason Cong}
\affiliation{%
  \institution{Computer Science Department, University of California, Los Angeles}
  \city{Los Angeles}
  \state{CA}
  \country{USA}
}
\email{cong@cs.ucla.edu}

\renewcommand{\shortauthors}{Sohrabizadeh and Yu, et al.}

\begin{abstract}
Adopting FPGA as an accelerator in datacenters is becoming mainstream for customized computing, but the fact that FPGAs are hard to program creates a steep learning curve for software programmers. Even with the help of high-level synthesis (HLS), accelerator designers still have to manually perform code reconstruction and cumbersome parameter tuning to achieve the optimal performance. While many learning models have been leveraged by existing work to automate the design of efficient accelerators, the unpredictability of modern HLS tools becomes a major obstacle for them to maintain high accuracy. To address this problem, we propose an automated DSE framework---{$\framework$}--- that leverages a bottleneck-guided coordinate optimizer to systematically find a better design point. {$\framework$} detects the bottleneck of the design in each step and focuses on high-impact parameters to overcome it. 
The experimental results show that {$\framework$} is able to identify the design point that achieves, on the geometric mean, {\speedupcpu}$\times$ speedup over one CPU core for Machsuite and Rodinia benchmarks. Compared to the manually optimized HLS vision kernels in Xilinx Vitis libraries, {$\framework$} can reduce their optimization pragmas by {\pragmareductionvitis} while achieving similar performance. With less than one optimization pragma per design on average, we are making progress towards democratizing customizable computing by enabling software programmers to design efficient FPGA accelerators.
\end{abstract}




\maketitle

\section{Introduction} \label{sec:intro}
Due to the rapid growth of datasets in recent years, the demand for scalable, high-performance computing continues to increase. However, the breakdown of Dennard's scaling~\cite{dennard74} has made the energy efficiency an important concern in datacenters, and has spawned exploration into using accelerators such as field-programmable gate arrays (FPGAs) to alleviate power consumption. For example, Microsoft has adopted CPU-FPGA systems in its datacenter to help accelerate the Bing search engine~\cite{catapult}; Amazon introduced the F1 instance~\cite{amazon-f1}, a compute instance equipped with FPGA boards, in its commercial Elastic Compute Cloud (EC2).

Although the interest in customized computing using FPGAs is growing, they are more difficult to program compared to CPUs and GPUs because the traditional register-transfer level (RTL) programming model is more like circuit design rather than software implementation. 
To improve the programmability, high-level synthesis (HLS)~\cite{cong11,zhang2008autopilot} has attracted a large amount of attention over the past decades. Currently, both FPGA vendors have their commercial HLS products---Xilinx Vitis~\cite{vitis-platform} and Intel FPGA SDK for OpenCL~\cite{intel-sdk}. With the help of HLS, one can program the FPGA more easily by controlling how the design should be synthesized from a high-level view. The main enabler of this feature is the ability to iteratively re-optimize the micro-architecture quickly just by inserting synthesis directives in the form of pragmas instead of re-writing the low-level behavioral description of the design. Because of the reduced code development cycle and the shorter turn-around times, HLS has been rapidly adopted by both academia and industry~\cite{hls4ml, lai2019heterocl,zohouri2018combined, sohrabizadeh2020end, andrade2017design, autosa}.
In fact, Code~\ref{code:cnn_hls_tranformed} shows an intuitive HLS C implementation of one forward path of a Convolutional Neural Network (CNN) on Xilinx FPGAs. 
Xilinx Vitis generates about 5800 lines of RTL kernel from Code~\ref{code:cnn_hls_tranformed} with the same functionality. As a result, it is much more convenient and productive for designers to evaluate and improve their designs in HLS C/C++.

Even though HLS is suitable for hardware experts to quickly implement an optimal design, it is not friendly for most of the general software designers who have limited FPGA domain knowledge. 
Since the hardware architecture inferred from a syntactic C implementation could be ambiguous, current commercial HLS tools usually generate architecture structures according to specific HLS C/C++ code patterns. 
As a result, even though it was shown in ~\cite{cong11} that the HLS tool is capable of generating FPGA designs with a performance as competitive as the one in RTL, not every C program gives a good performance and designers must manually reconstruct the HLS C/C++ kernel with specific code patterns and hardware specific pragmas to achieve high performance. 
As a matter of fact, the generated FPGA accelerator from Code~\ref{code:cnn_hls_tranformed} is 80$\times$ slower than a single-thread CPU. However, the optimized code (shown in Code~\ref{code:cnn_hls_tranformed_opt} in Appendix~\ref{a1_cnn})
is able to achieve around 7,041$\times$ speedup after we analyze and resolve several performance bottlenecks listed in Table~\ref{tbl:cnn_analysis} by applying code transformations and inserting 28 pragmas. 

\begin{lstlisting}[language=C,caption=CNN HLS C Code Snippet\label{code:cnn_hls_tranformed},
    float,floatplacement=H]
// Skip const variable initialization and macro definitions for brevity
void CnnKernel(const float* input $\circled{\small 1}$, const float* weight$\circled{\small 1}$, 
               const float* bias $\circled{\small 1}$, float* output $\circled{\small 1}$) {

  float C[ParallelOut][ImSize][ImSize];
  for (int i = 0; i < NumOut / ParallelOut; ++i) { $\circled{\small 4}$ 
    // Initialization
    for (int h = 0; h < ImSize; ++h) {
      for (int w = 0; w < ImSize; ++w) {
        for (int po = 0; po < ParallelOut; po++)
            C[po][h][w] = bias[(i << shift) + po]; } }
    // Convolution
    for (int j = 0; j < NumIn; ++j) { $\circled{\small 5}$
      for (int h = 0; h < ImSize; ++h) { $\circled{\small 5}$
        for (int w = 0; w < ImSize; ++w) { $\circled{\small 5}$
          for (int po = 0; po < ParallelOut; po++) { $\circled{\small 5}$
            for (int p = 0; p < kKernel; ++p) { $\circled{\small 5}$
              for (int q = 0; q < kKernel; ++q) $\circled{\small 5}$
                C[po][h][w] += weight(i, po, j, p, q) * input(j,h + p,w + q); $\circled{\small 2}$ $\circled{\small 3}$ } } } } }
    // ReLU + Max pooling
    for (int h = 0; h < OutImSize; ++h) { $\circled{\small 5}$
      for (int w = 0; w < OutImSize; ++w) { $\circled{\small 5}$
        for (int po = 0; po < ParallelOut; po++) { $\circled{\small 5}$
          output(i,h,w) = max(0.f, C, po, h, w);  } } } } }
\end{lstlisting}
\begin{table}[!th]
\centering
\caption{Analysis of Poor Performance in Code~\ref{code:cnn_hls_tranformed}}
\label{tbl:cnn_analysis}
\begin{tabularx}{\linewidth}{|c|>{\hsize=.29\hsize}X|>{\hsize=.68\hsize}X|}
\hline
 & Reason & Required Code Changes for Higher Performance\\ \hline\hline
$\circled{\small 1}$ & Low bandwidth util. & Manually apply memory coalescing using HLS built-in type \texttt{ap\_int}. \\ \hline
$\circled{\small 2}$ & High access latency to global memory & Manually allocate local buffer and use memcpy to enable memory burst. \\ \hline
$\circled{\small 3}$ & Does not hide communication latency & Manually create load/compute/store functions and double buffering. \\ \hline
$\circled{\small 4}$ & Lack of parallelism & Manually create parallel  coarse-grained processing elements by wrapping the inner loops as a function and setting proper array partition factors. \\ \hline
$\circled{\small 5}$ & Sequential execution & Apply \texttt{\#pragma HLS pipeline} and \texttt{\#pragma HLS unroll} with proper array partition factors for each processing element. \\ \hline
\end{tabularx}
\end{table}

It turns out that the bottlenecks presented in Table~\ref{tbl:cnn_analysis} occur for most C/C++ programs developed by software programmers, and similar optimizations have to be repeated for each new application, which makes HLS C/C++ design not scalable.
In general, there are three levels of optimization that one needs to employ to get to a high-performance FPGA design.
The level one is for increasing the data reuse or reducing/removing the data dependency by \textit{loop} transformations, which is common in CPU performance optimizations as well (e.g. for cache locality); therefore, it is well accepted by software programmers and we expect them to apply such transformations manually without any problems.
The second level is required to enable repetitive architectural optimizations that most of the designs benefit from, such as memory burst and memory coalescing, as mentioned in reasons 1-2 in Table~\ref{tbl:cnn_analysis}. Fortunately, the recently developed Merlin Compiler \footnote{The Merlin Compiler will be open-sourced in the near future after passing Xilinx's legal review.}~\cite{merlin, merlin_islped, fcs}  from Falcon Computing Solutions~\cite{fcs}, which was acquired by Xilinx in late 2020~\cite{acquisition}, can automatically take care of this kind of code transformations.

The final and the most critical level deals with FPGA-specific architectural optimizations, detailed in reasons 3-5 in Table~\ref{tbl:cnn_analysis}, that vary from application to application. Although the Merlin Compiler also helps alleviate this problem to some extent by introducing a few high-level optimization pragmas and applying source-to-source code transformation to enable them, these optimizations are much more difficult for software programmers to learn and apply effectively. More specifically, choosing the right part of the program to optimize, deciding the type of optimization and the pragmas to apply for enabling it, and tuning the pragma to get to the design with the highest quality complicate this level. 

Apparently, the requirement of mastering all three levels of optimizations makes the bar for general software programmers to use FPGA extremely high.
Hence, general software programmers will lean towards other popular accelerators such as power-consuming GPUs or high-cost ASICs with less considerations over FPGAs.
These obstacles consequently result in huge barriers in the adoptions of FPGA in datacenters, the expansion of the FPGA user community, and the advances of FPGA techology. 
One possible solution is to apply an automated micro-architecture optimization. 
Thus, everyone with decent knowledge of programming is able to try customized computing with minimum effort. In order to free accelerator designers from the iterations of HLS design improvement, automated design space exploration (DSE) for HLS attracts more and more attention. 
However, existing DSE methods face the following challenges:

\textbf{Challenge 1: The large solution space:}
 The solution space grows exponentially by the number of candidate pragmas. In fact, only applying pipeline, unroll, and array partition pragmas to Code~\ref{code:cnn_hls_tranformed} produces $10^{20}$ design points. This huge number of combinations creates a serious impediment to exploring the whole design space.

\textbf{Challenge 2: Non-monotonic effect of design parameters on performance/area:}
As pointed out by Nigam, et al.~\cite{nigam2020predictable}, we cannot assume that an individual design parameter will affect the performance/area in a smooth and/or monotonic way.

\textbf{Challenge 3: Correlation of different characteristics of a design:}
When different pragmas are employed together in a design, they do not affect only one characteristic of a design. 
We will use the convolution part of the Code~\ref{code:cnn_hls_tranformed} as an example. If we apply fine-grained (\textit{fg}) pipeline to \texttt{w} loop and parallelize the loop with a factor of 2, it results in a loop with initiation interval (II) of 2 synthesized by Vivado HLS~\cite{hls}. However, when we change the parallel factor to 4, the HLS tool increases the II to 3 to optimize resource consumption by reusing some of the logic units instead of doubling the resource utilization. The analytical models usually fail to capture these cases. 
Furthermore, pipelining the \texttt{j} loop is part of the best design configuration. However, it does not improve the performance until after the \textit{fg} pipelining is applied on the \texttt{w} loop. It suggests that the order of applying the pragmas is crucial in designing the explorer.

\textbf{Challenge 4: Implementation disparity of HLS tools:}
The HLS tools from different vendors employ different implementation strategies. Even within the same vendor, the optimization and implementation rules keep changing across different versions. 
For example, the past Xilinx SDAccel versions consistently utilize \textit{registers} to implement array partitions with small sizes to save BRAMs. However, the latest ones use \textit{dual-port BRAMs} for implementation to support two reads in one cycle for achieving full pipelining, or II = 1, even if the array size is small. Such implementation details are hard to capture and maintain in analytical models and make it difficult to port an analytical model built on a specific tool to the other.

\textbf{Challenge 5: Long synthesis time of HLS tools: }
HLS tools usually take 5-30 minutes to generate RTL and estimate the performance---and even longer if the design has a high performance. This emphasizes the need for a DSE that can find the Pareto-optimal design points in fewer iterations.

In this paper, as our first step to lowering the bar for general software programmers to make the FPGA programming universally accessible, we focus on automating the final level of optimization.
To solve the challenges 2 to 4 mentioned above, instead of developing an analytical model, we treat the HLS tool as a black-box.
Challenges 1 and 5 imply that we need to explore the solution space intelligently. 
For that, we first apply the coordinate descent with the finite difference method to guide the explorer. However, we show that the general application-oblivious approaches fail to perform well for the HLS DSE problem. As a result, we present the {$\framework$} \footnote{All the codes will be open-sourced after the paper is accepted.} framework that adapts a bottleneck-guided coordinate optimizer to systematically search for better configurations. 
We incorporate a flexible list-comprehension syntax to represent a grid design space with all invalid points marked.
In addition, we also partition the design space systematically to address the local optimum problem caused by Challenge 2.

In summary, this paper makes the following contributions:
\begin{itemize}
    \item We propose two strategies to guide DSE. One adapts the commonly used coordinate descent with the finite difference method and the other exploits a bottleneck-guided coordinate optimizer.
    \item We incorporate list-comprehension to represent a smooth, grid design space with all invalid points marked.
   \item We develop the {$\framework$} framework on top of the Merlin Compiler to automatically perform DSE using the bottleneck optimizer 
    to systematically close in on high-QoR design points.
    \item To the best of our knowledge, we are the first ones to evaluate our tool using the Xilinx optimized vision library~\cite{vitis}. Evaluation results indicate that $\framework$ is able to achieve the same performance, yet with {\pragmareductionvitis} reduction of their optimization pragmas resulting in less than one required optimization pragma per kernel, on the geometric mean.
    \item We evaluate {$\framework$} on 11 computational kernels from Machsuite~\cite{machsuite} and Rodinia~\cite{che2009rodinia} benchmarks and one convolution layer of Alexnet~\cite{alexnet}, 
    showing that we are able to achieve, on the geometric mean, {\speedupcpu}$\times$ speedup over a single-thread CPU---only a 7\% performance gap compared to manual designs. 
\end{itemize}
\vspace{-0.2cm}
\section{Problem Formulation} \label{sec:bg}

Our goal is to expedite the hardware design by automating its exploration process. 
In general, there are two types of pragmas (using Vivado HLS as an example) that are applied to a program. One type is the \textit{non-optimization} pragmas, 
which are relatively easy for software programmers to learn and apply. The other type is \textit{optimization} pragmas, including \texttt{PIPELINE} and \texttt{UNROLL} pragmas. These pragmas require knowledge of FPGA devices and micro-architecture optimization experience, which are usually much more challenging for a software programmer to learn and master as explained in Section~\ref{sec:intro}. 
The goal of this research is to minimize or eliminate the need to apply optimization pragmas \textit{manually} and let $\framework$ insert them \textit{automatically}. More formally, we formulate the HLS DSE problem as the following:

\noindent\textit{\textbf{Problem 1: Identify Design Space.}} Given a C program $\mathcal{P}$ as the FPGA accelerator kernel, construct a design space $\mathbb{R}^{K}_{\mathcal{P}}$ with $K$ parameters that contains possible combinations of HLS pragmas for $\mathcal{P}$ as design configurations.

\noindent\textit{\textbf{Problem 2: Find the Optimal Configuration.}} Given a C program $\mathcal{P}$, we would like to insert a minimal number of optimization pragmas manually to get a new program $\mathcal{P'}$ as the FPGA accelerator kernel along with its design space set $\mathbb{R}^{K}_{\mathcal{P'}}$ which is identified in Problem 1, and we let the DSE tool insert the rest of the pragmas automatically. More specifically, having a vendor HLS tool $\mathbf{H}$ that estimates the execution cycle $Cycle(\mathbf{H},\mathcal{P'})$ and the resource utilization $Util(\mathbf{H},\mathcal{P'})$ of the given $\mathcal{P'}$ as a black-box evaluation function, the DSE must find a configuration $\theta \in \mathbb{R}^{K}_{\mathcal{P'}}$ in a given search time limit so that the generated design $\mathcal{P'}(\theta)$ with $\theta$ can fit in the FPGA and the execution cycle is minimized. Formally, our objective is:
\begin{align}
\underset{\theta}{\min}\ Cycle(\mathbf{H}, \mathcal{P'}(\theta))
\end{align}
\vspace{-0.3cm}
\noindent subject to
\begin{align}
\begin{split}
\theta &\in \mathbb{R}^{K}_{\mathcal{P'}}\\
\forall u &\in Util(\mathbf{H}, \mathcal{P'}(\theta)), u < T_u\
\end{split}
\end{align}

\noindent where $u$ is the utilization of one of the FPGA on-chip resources and $T_u$ is a user-available resource threshold on FPGAs. We set all $T_u$ to 0.8, an empirical threshold, in our experiments. Beyond 0.8, the design will suffer from high clock frequency degradation due to the difficulty in placement and routing. In addition, the rest of the resources are left for the interface logic of the vendor HLS tool. 

Note that we introduce two optimization objectives; one minimizes the optimization pragmas that has to be inserted manually to obtain $\mathcal{P'}$, and the other maximizes the performance of $\mathcal{P'}$ using $\framework$ by applying pragmas automatically. Obviously, there is a trade-off between the two. An expert designer can always get an optimized micro-architecture to achieve the best performance by inserting enough HLS optimization pragmas. However, it is time-consuming and not feasible for software programmers with little or no FPGA design experience. In our evaluation, our goal is to match the performance of well-designed HLS library code (typically written by experts) yet insert much fewer optimization pragmas \textit{manually}. Indeed, our experimental results in Section~\ref{sec:eval} show that we can achieve this with {\pragmareductionvitis} pragma reduction on the geometric mean, requiring less than 1 optimization pragma per kernel.

\vspace{-0.2cm}
\section{Related Work} \label{sec:relWork}
There are a number of previous works that propose an automated framework to explore the HLS design space, and they can be summarized in two categories: model-based and model-free techniques. 
\subsection{Model-based Techniques}
The studies in this category build an analytical model for evaluating the quality of each explored design point by estimating its performance and resource utilization. The authors in~\cite{wang2017flexcl, comba, linanalyzer} build the dependence graph of the target application and utilize traditional graph analysis techniques along with predictive models to search for the best design. Although this approach can quickly search through the design space, it is inaccurate and it is difficult to maintain the model and port it to other HLS tools as explained in Challenge 4 of Section~\ref{sec:intro}. Zhong, et al.~\cite{zhong2014design} develops a simple analytical model for performance and area estimation. However, they assume that the performance/area changes monotonically by modifying an individual design parameter, which is not a valid assumption as we explained in Challenge 2 of Section~\ref{sec:intro}. To increase the accuracy of the estimation model, a number of other studies restrict the target application to those that have a well-defined accelerator micro-architecture template ~\cite{chi2018soda, cong2018polysa, autoaccel, reggiani2019pareto, sohrabizadeh2020end, zacharopoulos2019compiler}, a specific application~\cite{xu2020autodnnchip, zheng2020flextensor}, or a particular computation pattern~\cite{choi2018hls,koeplinger2016automatic,prabhakar2016generating}; hence, they lose generality. 

To the same end, there are other studies that build the predictive model using learning algorithms. They train a model by iteratively synthesizing a set of sample designs and updating the model until it gets to the desired accuracy. Later on, they use the trained model for estimating the quality of design instead of invocations of the HLS tool. To learn the behavior of the HLS tool, these works adapt supervised learning algorithms to better capture uncertainty of HLS tools~\cite{koeplinger2016automatic,liu2013learning,  liu2019accelerating,schafer2012machine,xydis2015spirit,zhong2017design}. While this technique increases the accuracy of the model, it is still hard to port the model to another HLS tool in a different vendor or version. Often by changing the HLS tool or the target FPGA, new samples should be collected which can be an expensive step. After that, for each of them, a new model should be trained to include the new dataset.

\subsection{Model-free Approaches}
To avoid dealing with the uncertainty of HLS tools, in this category, the studies treat the HLS tool as a black box. Instead of learning a predictive model, they invoke HLS every time to evaluate the quality of the design. To guide the search, they either exploit general application-oblivious heuristics (e.g., simulated annealing~\cite{mahapatra2014machine} and genetic algorithm~\cite{schafer2017parallel}) or they develop their own heuristics~\cite{ferretti2018cluster, ferretti2018lattice, schafer2012divide}. S2FA~\cite{s2fa} 
employ multi-armed bandit~\cite{Fialho2010} to combine a set of heuristic algorithms including uniform greedy mutation, differential evolution genetic algorithm, particle swarm optimization, and simulated annealing.
However, as we will present in Section~\ref{sec:learn}, general hyper-heuristic approaches are unreliable for finding the high quality of result (QoR) design configuration. Moreover, the authors in~\cite{ferretti2018cluster, ferretti2018lattice} claim that Pareto-optimal design points cluster together. They exploit an initial sampling to build the first approximation of the Pareto frontier and require local searches to explore other candidates. However, the cost of initial sampling is not scalable when the design space is tremendously large
(e.g., the scale of $10^{10}$ to $10^{30}$), as the ones we have enumerated in this paper are. Sun et, al~\cite{date21} adapt a (Gaussian process) GP-based Bayesian optimization (BO) algorithm  to explore the solution space. At each iteration, it improves a surrogate model to mimic the HLS tool, by sampling the design space. Again, as the search space grows, it will require more samples to build a good surrogate model which can limit the scalability. Moreover, the computation of a GP-based BO can be seen to be cubic in the total number of samples (in addition to the time to evaluate the sampled point using the HLS tool), as it wants to calculate the inversion of a dense covariance matrix at each step~\cite{snoek2015scalable} which can further limit the scalability of the approach.
\section{The {\framework } Framework} \label{sec:framework}
To reduce the size of the design space, we build our DSE on top of the Merlin Compiler~\cite{merlin, merlin_islped}. Section~\ref{sec:bg_fpga} reviews the Merlin Compiler and justifies our choice. Then, we present an overview of $\framework$ in Section~\ref{sec:framework_overview}.

\subsection{Merlin Compiler and Design Space Definition} \label{sec:bg_fpga}
\begin{table}[!htb]
\centering
\caption{Merlin Pragmas with Architecture Structures}
\label{tbl:merlin_pragmas}
\begin{tabular}{ccc}
\hline
\multicolumn{1}{|c|}{Keyword}                   & \multicolumn{1}{c|}{Available Options}             & \multicolumn{1}{c|}{Architecture Structure} \\ \hline\hline
\multicolumn{1}{|c|}{parallel}                  & \multicolumn{1}{c|}{factor=\textless{}int\textgreater{}}                    & \multicolumn{1}{c|}{CG \& FG parallelism}   \\ \hline
\multicolumn{1}{|c|}{\multirow{2}{*}{pipeline}} & \multicolumn{1}{c|}{mode=cg}                       & \multicolumn{1}{c|}{CG pipeline}            \\ \cline{2-3} 
\multicolumn{1}{|c|}{}                          & \multicolumn{1}{c|}{mode=fg}                       & \multicolumn{1}{c|}{FG pipeline}            \\ \hline
\multicolumn{1}{|c|}{tiling}                  & \multicolumn{1}{c|}{factor=\textless{}int\textgreater{}}                    & \multicolumn{1}{c|}{Loop Tiling}   \\ \hline
\multicolumn{3}{c}{CG: Coarse-grained; FG: Fine-grained}
\end{tabular}
\end{table}
The Merlin Compiler~\cite{merlin, merlin_islped} was developed to raise the abstraction level in FPGA programming by introducing a reduced set of high-level optimization directives and generating the HLS code according to them automatically. It uses a simple programming model similar to OpenMP~\cite{dagum1998openmp}, which is commonly used for multi-core CPU programming. Like in OpenMP, it defines a small set of compiler directives in the form of pragmas for optimizing the design. Table~\ref{tbl:merlin_pragmas} lists the Merlin pragmas with architecture structures. Note that the \texttt{fg} option in the fine-grained pipeline mode refers to the code transformation that tries to apply fine-grained pipelining to a loop nest by fully unrolling all its sub-loops; whereas, the \texttt{cg} option in the coarse-grained pipelining transforms the code to enable double buffering. Based on these user-specified pragmas, the Merlin Compiler performs source-to-source code transformation and automatically generates the related HLS pragmas such as \texttt{PIPELINE}, \texttt{UNROLL}, and \texttt{ARRAY\_PARTITION} to apply the corresponding architecture optimization.

To reduce the size of the solution space, we chose to utilize the Merlin Compiler  
as the backend of our tool. 
 Since the number of pragmas required by the Merlin Compiler is much smaller (as it performs source level code reconstruction and generates most of the HLS required pragmas), it defines a more compact design space, which makes it a better fit for developing a DSE as shown in~\cite{autoaccel,s2fa}. For instance, Code~\ref{code:merlin-cnn} shows the CNN kernel with Merlin pragmas. With inserting only four lines of pragmas and no further \textit{manual} code transformation, the Merlin Compiler is able to transform Code~\ref{code:merlin-cnn} to a high-performance HLS kernel with the same performance as the manually optimized design written in HLS C which has 28 pragmas as mentioned in Section~\ref{sec:intro}.
 
The Merlin Compiler, by default, applies code transformations to address the bottlenecks 1 and 2 listed in Table~\ref{tbl:cnn_analysis} and provides high-level optimization pragmas for the rest of them. For example, instead of rewriting Code~\ref{code:cnn_hls_tranformed} to test whether double buffering would help the performance as described in reason 3 in Table~\ref{tbl:cnn_analysis}, we just need to use the \texttt{cg PIPELINE} pragma and the Merlin Compiler will rewrite the code to satisfy it. 
As a result, our focus in this work is on finding the best location of each of these high-level pragmas and tuning them, automatically; hence, we can address reasons 3-5 in Table~\ref{tbl:cnn_analysis} as well by enabling the architectural optimizations along with the best pipelining and parallelization attributes.

\begin{lstlisting}[caption=CNN Code Snippet in Merlin C\label{code:merlin-cnn},
    float,floatplacement=H]
void CnnKernel(
	const float input[NumIn][InImSize][InImSize],
	const float weight[NumOut][NumIn][kKernel][kKernel], 
	const float bias[NumOut], float output[NumOut][OutImSize][OutImSize]) {

  float C[ParallelOut][ImSize][ImSize];
  for (int i = 0; i < NumOut/ParallelOut; i++) {
    // Initialization
	  for (int h = 0; h < ImSize; ++h) {
#pragma ACCEL parallel factor=4
	    for (int w = 0; w < ImSize; ++w){
	      for (int po = 0; po < ParallelOut; po++)
		        C[po][h][w] = 0.f; } }
    // Convolution
#pragma ACCEL pipeline
	  for (int j = 0; j < NumIn; ++j) {
	    for (int h = 0; h < ImSize; ++h) {
#pragma ACCEL parallel factor=4
#pragma ACCEL pipeline FLATTEN 
	      for (int w = 0; w < ImSize; ++w) {
		      for (int po = 0; po < ParallelOut; po++){
		        float tmp = 0;
			      for (int p = 0; p < kKernel; ++p) {	 
			        for (int q = 0; q < kKernel; ++q){
				        tmp += ... } }
			      C[po][h][w] += tmp; } } } }
    // Skip ReLU + Max Pooling for brevity
} } 
\end{lstlisting}


As a result, our solution to Problem~1 is defined as in Table~\ref{tbl:ds}. We identify the design space for each kernel by analyzing the kernel abstract syntax tree (AST) to gather loop trip-counts, available bit-widths, etc. The rules we enforce in building this design space are listed in Section~\ref{sec:gradient_ds}

\begin{table}[!htb]
\centering
\caption{Design Space Building on Merlin Pragmas}
\label{tbl:ds}
\begin{tabular}{|c|l|}
\hline
\rule{0pt}{2ex} Factor & Design Space (Values) \\[0.5ex]  \hline\hline
\rule{0pt}{2ex} CG-loop parallel & $\left\{u \mid 1 < u <= TC(L), u.c = TC(L), c \in \mathbb{Z} \right\}$ \\ [0.5ex] \hline
\rule{0pt}{2ex} FG-loop parallel & {\scriptsize$\left\{u \mid 
\begin{cases}
1 < u < TC(L), u.c = TC(L), c \in \mathbb{Z}, & \mbox{TC(L) > 16} \\
u = TC(L), & \mbox{otherwise}
\end{cases}\right\}$} \\ [0.5ex] \hline
\rule{0pt}{2ex} CG-loop pipeline & $\left\{p \mid p \in \left\{off,cg,fg\right\}\right\}$ \\ [0.5ex] \hline
\rule{0pt}{2ex} FG-loop pipeline & $\left\{p \mid p = fg\right\}$ \\ [0.5ex] \hline
\rule{0pt}{2ex} loop tiling & $\left\{t \mid 1 < t < TC(L), t.c = TC(L), c \in \mathbb{Z}\right\}$ \\ [0.5ex] \hline
\multicolumn{2}{l}{CG: Coarse-grained; FG: Fine-grained; TC: Loop trip-count}
\end{tabular}
\end{table}
Now that we have defined the design space in Table~\ref{tbl:ds} for \textit{\textbf{Problem 1}}, we focus on \textit{\textbf{Problem 2}} in the remainder of this paper. Although to some extents, Merlin pragmas alleviate the manual code reconstruction overhead, a designer still has to manually search for the best option for each pragma, including position, type, and factors. In fact, choices for the CNN design in Code~\ref{code:cnn_hls_tranformed} contain four DRAM buffers and thirteen loops, which result in $\sim10^{16}$ design configurations. The large design space motivates us to develop an efficient approach to find the best configuration.

\subsection{Framework Overview}
\label{sec:framework_overview}
\begin{figure}[!tbh]
	\centering
	\includegraphics[scale=0.5]{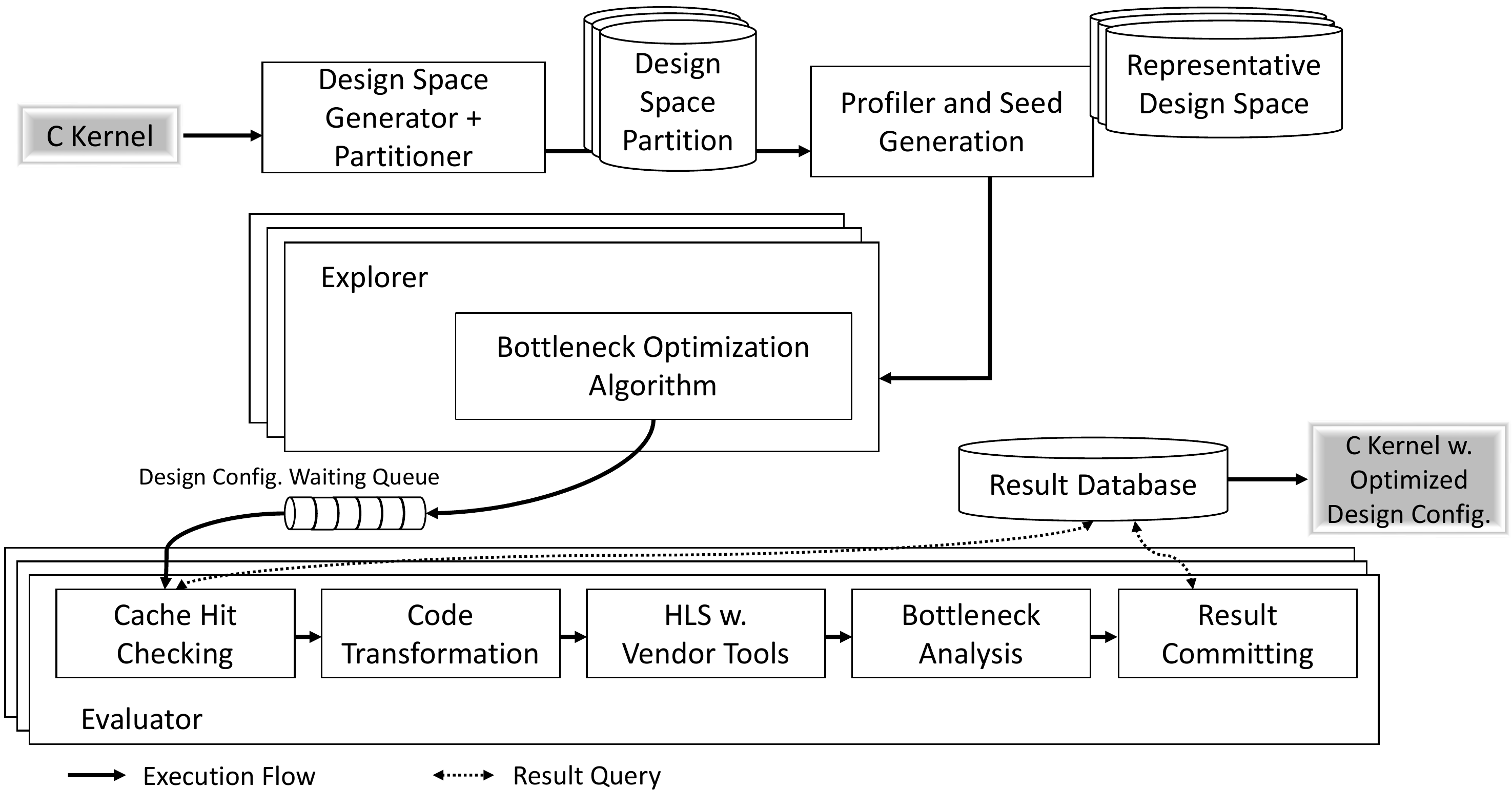} 
	\caption{The {$\framework$} Framework Overview}
	\label{fig:framework}
\end{figure}

We develop and implement {$\framework$}, a push-button framework, as depicted in Fig.~\ref{fig:framework} based on the strategies explained in Section~\ref{sec:gradient}. The framework first automatically builds a design space by analyzing the kernel AST according to the rules and the syntax described in Section~\ref{sec:gradient_ds}. Then, it profiles and selects representative partitions using K-Means as mentioned in Section~\ref{sec:gradient_partition}. For each partition, {$\framework$} explorer performs DSE using the proposed bottleneck-based coordinate strategy in Section~\ref{sec:bottleneck_algo} and the parameter ordering explained in Section~\ref{sec:param-order}. The explorer can be tuned to evaluate the quality of design points based on different targets such as performance, resource, or finite difference (Eq.~\ref{eq:fd}). 
When the explorer finishes exploring a partition, it stores the best configuration found by that partition and reallocates the working threads to other partitions to keep the resource utilization high. Finally, when all partitions are finished, {$\framework$} outputs the design configuration with the best QoR among all partitions.

\section{{\framework} Methodology} \label{sec:gradient}

In this section, we first examine the efficiency of application-oblivious heuristics, which were considered in our initial study, in Section~\ref{sec:learn}. As we will discuss, the main drawback of these heuristics for the HLS DSE problem is the fact that they do not have any knowledge of the semantics of the program parameter. This problem can potentially linger the DSE process since the explorer may waste a lot of time on parameters with no impact on the results at that stage of optimization.
As a result, in Section~\ref{sec:bottleneck_algo}, we present a bottleneck-guided coordinate optimizer that can mimic an expert's optimization method and generate high-QoR design points in fewer number of iterations. 
We propose several optimizations in Sections~\ref{sec:param-order} to~\ref{sec:gradient_partition} to further improve the performance of our framework.

\subsection{\textbf{Application-oblivious Heuristics}} \label{sec:learn}
In our prior work on DSE, S2FA~\cite{s2fa}, we adapted a popular search engine called OpenTuner~\cite{opentuner}. OpenTuner leverages the multi-armed bandit (MAB) approach~\cite{Fialho2010} to assemble multiple meta-heuristic algorithms - including uniform greedy mutation, differential evolution genetic algorithm, particle swarm optimization, and simulated annealing - for high generalization. At each iteration, the MAB selects the meta-heuristic with the highest credit and updates the credit of the selected meta-heuristic based on the QoR, which means the meta-heuristic that can efficiently find high-quality design points will be rewarded and activated more frequently by the MAB, and vice versa. Due to its extensibility, OpenTuner has been adapted to perform DSE for hardware design optimization~\cite{datuner}. 
 
\begin{figure}[!tbh]
	\centering
	\includegraphics[scale=0.6]{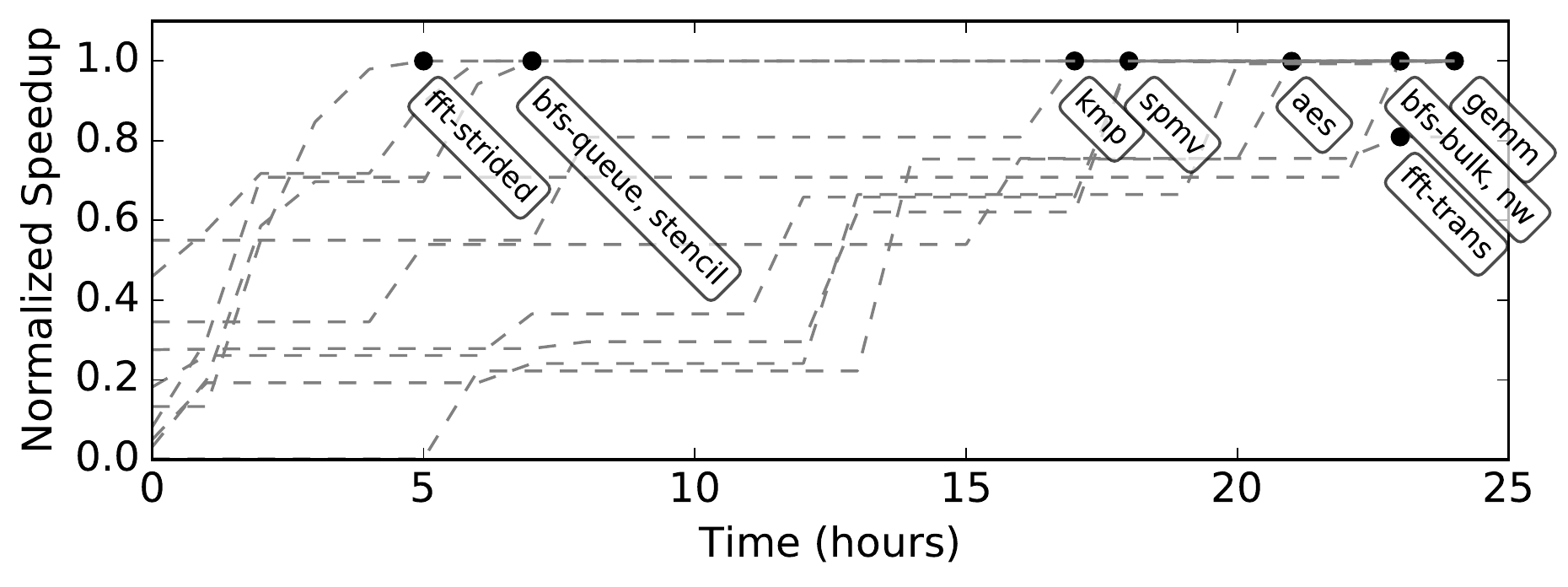} 
	\caption{Speedup Over the Manual Design Using S2FA~\cite{s2fa}}
	\label{fig:exp_trend}
\end{figure}

S2FA also employs the Merlin Compiler as its backend and further applies more strategies to improve the OpenTuner efficiency when performing DSE for HLS. We use S2FA to perform the DSE for 24 hours and depict the speedup of our benchmark cases compared to the corresponding manual design over time in Fig.~\ref{fig:exp_trend}. The black dot indicates the time that the S2FA finds the overall best design point. We can see that S2FA requires on average 16.8 hours to find the best solution. We further analyze the exploration process and find that most designs have an obvious performance bottleneck (e.g., low utilization of global memory bandwidth, insufficient parallel factors, etc.), which usually dominates more than half of the overall execution cycle and is controlled by only one or two design parameters (pragmas). In this situation, the performance gain of tuning other parameters is often very limited but it is hard for the problem-independent searching algorithm to learn that. In fact, it needs many iterations to identify the key parameter and tune it to resolve the performance bottleneck. After that, it has to spend a large number of iterations again to find the next key parameter. This phenomenon motivates us to develop a new search algorithm that is guaranteed to optimize the key parameter (high-impact parameter) prior to others.

Coordinate descent is another well-known iterative optimization algorithm for finding a locally minimum point. It is based on the idea that one can minimize a multi-variable function by minimizing it along one direction at a time and solving single-variable optimization problems. 
At each iteration, we generate a set of candidates, $\Theta_{cand}$, as the input to the algorithm. Each candidate is generated by advancing the value of each parameter in the current configuration by one step. Formally, the $c$-th candidate generated from design point $\theta_i$ is:
\begin{align} \label{eq:candidate}
    \theta_i^c = [p_0, p_1, ..., p_c + 1, ..., p_K]
\end{align}
\noindent where $K$ is the total number of parameters, $p_c$ is the value of $c$-th parameter in $\theta_i$, $p_c + 1$ denotes the next value of this parameter (the next numeric factor for \texttt{PARALLEL} and \texttt{TILING} pragma and the next mode of pipelining for \texttt{PIPELINE} pragma). Accordingly, we will generate $K$ candidates at each iteration, which means we run HLS $K$ times to determine the next configuration as follows:
\begin{align}
    \theta_{i+1} = \underset{\theta_i^c \in \Theta_{cand}}{argmin} g(\theta_i^c, \theta_i)
\end{align}

We leverage the finite difference method to approximate the coordinate value by treating the HLS tool as a black-box. That is, given a candidate configuration $\theta_j$ deviated from the current configuration $\theta_i$, the coordinate value is defined as:
\begin{align} \label{eq:fd}
    g(\theta_j, \theta_i) \sim \frac{Cycle(\mathbf{H}, \mathcal{P}(\theta_j)) - Cycle(\mathbf{H}, \mathcal{P}(\theta_i))}{Util(\mathbf{H}, \mathcal{P}(\theta_j)) - Util(\mathbf{H}, \mathcal{P}(\theta_i))}
\end{align}

\noindent We calculate $Util(\mathbf{H}, \mathcal{P}(\theta))$ by taking into account all the different types of resources using the following formula:
\begin{align} \label{eq:util}
    Util(\mathbf{H}, \mathcal{P}(\theta)) = \sum_{u} 2^{\frac{1}{1-u}}
\end{align}
where $u$ is the utilization of one of the FPGA resources. We use exponential function to penalize the over-utilization of FPGA more seriously. \noindent Note that Eq.~\ref{eq:fd} considers not only performance gain but also resource efficiency, so it could reduce the possibility of being trapped in a local optimum. For example, we may reduce 10\% execution cycle by spending 30\% more area if we increase the parallel factor of a loop (configuration $\theta_1$); we can also reduce 5\% execution cycle by spending 10\% more area if we enlarge the bit-width of a certain buffer (configuration $\theta_2$). Although $\theta_1$ seems better in terms of the execution cycle, it may be more easily trapped by a locally optimal point because it has a relatively limited resource left to be further improved. On the other hand, the finite difference values for the two configurations are $g(\theta_1, \theta_0)=\frac{-10\%}{30\%}=-0.3$ and $g(\theta_2, \theta_0)=\frac{-5\%}{10\%}=-0.5$, so the system prioritizes the second configuration for a better long-term performance.

By leveraging the coordinate descent with a finite difference method, we expect to find a better design point every $K$ HLS runs. Unfortunately, as mentioned in Challenge 2 of Section~\ref{sec:intro}, the performance trend is not always smooth, so the coordinate process can easily be trapped by a low-quality locally optimal design point. 
Actually, this approach only achieves 2.8$\times$ speedup, on the geometric mean, for our  MachSuite~\cite{machsuite} and Rodinia~\cite{che2009rodinia} (MR) benchmarks, which is even worse than the results from S2FA. 

Moreover, the efficiency of using the coordinate-based approach for DSE is limited by the number of parameters. More specifically, at each iteration, we need to evaluate $K$ design points, where $K$ is the total number of tuning parameters, to determine the next step. On the other hand, in most cases, only a few of the $K$ tuning parameters have a high impact on the performance, so we should evaluate only the $K'$ impactful parameters at each iteration where $K' < K$. For instance, design space generator will instrument Code~\ref{code:cnn_hls_tranformed} with 27 pragmas based on the rules explained in Section~\ref{sec:gradient_ds} and the coordinate-based approach proposed in this section needs to assess the quality of 27 new designs in each iteration. However, in the early iterations the convolution part takes more than 85\% of the total cycle counts of the kernel. As a result, changing the pragmas outside of this part will have insignificant effect on the performance; hence, it is wasteful to explore them at this stage.

\subsection{\textbf{{\framework} Exploring Strategy - Bottleneck-guided Coordinate Optimizer}}
\label{sec:bottleneck_algo}
\begin{figure}[!tbh]
	\centering
	\includegraphics[width=\linewidth]{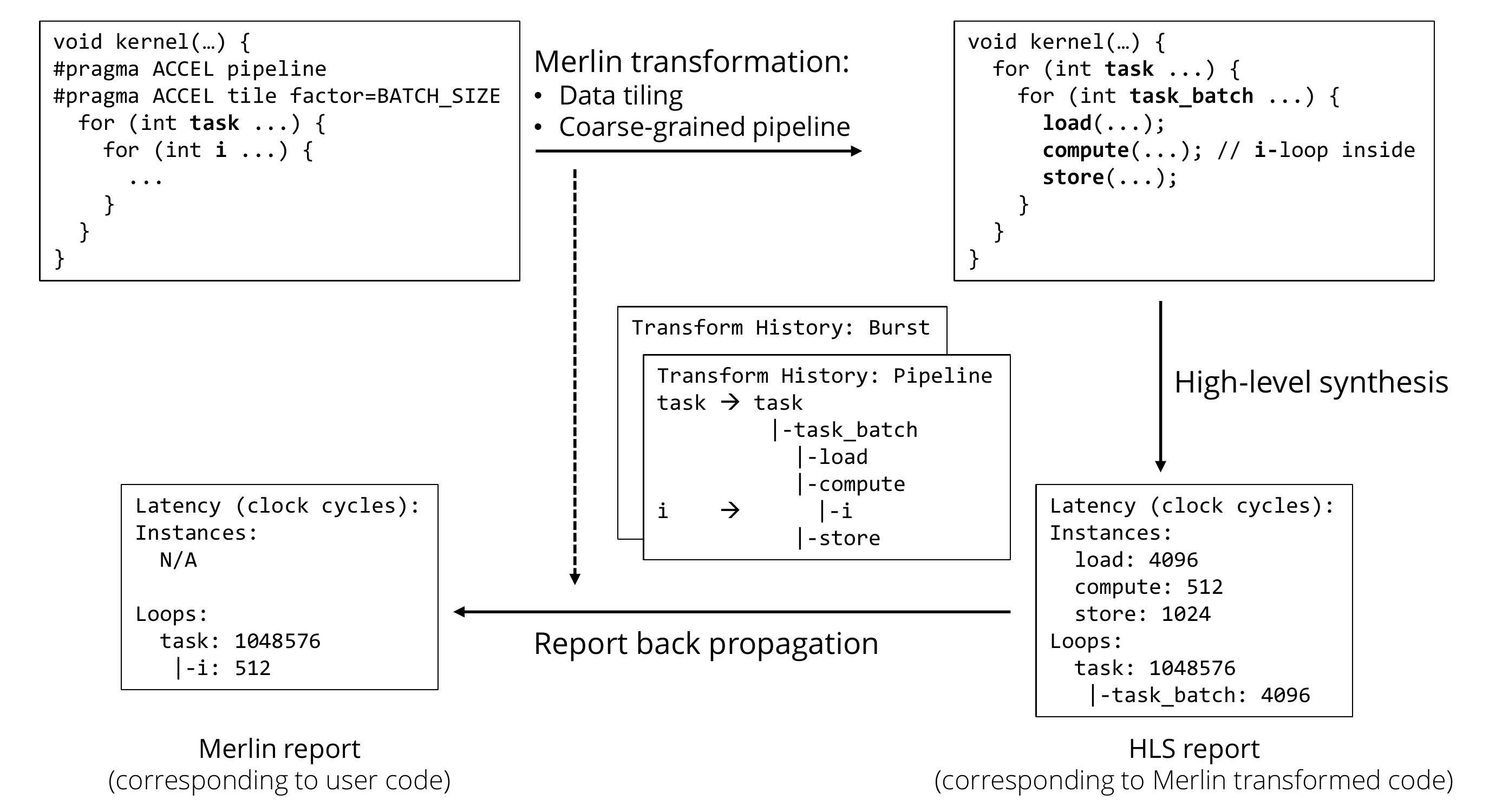} 
	\caption{The Cycle Propagation in Merlin Compiler}
	\label{fig:merlin-report}
\end{figure}

Two main inefficiencies of the approaches reviewed in the previous section are 1) they must evaluate many design points to identify the performance bottleneck, 2) they have no knowledge of the semantics of the parameters, so they have no way to differentiate them and prioritize the important ones. 
Identifying the key parameters is not straightforward. Although HLS report may provide the cycle breakdown for the loop and function statements, it is hard to map them to tuning parameters due to the applications of several code transformations applied by the Merlin Compiler. Fortunately, the Merlin Compiler includes a feature that transmits the performance breakdown reported by the HLS tool to the user input code, allowing us to identify the performance bottleneck by traversing the Merlin report and mapping the bottleneck statement to one or few tuning parameters.

Fig.~\ref{fig:merlin-report} illustrates how the Merlin Compiler generates its report of the cycle breakdown. When performing code transformation, the Merlin Compiler records the code change step by step so that it is able to propagate the latency estimated by the HLS tool back to the user input code. In this example, the \texttt{i} loop corresponds to the \texttt{compute} unit in the transformed code, so the latency of this unit is assigned to it. Note that the latency of all \texttt{load}, \texttt{compute}, and \texttt{store} units are included in the \texttt{task\_batch} loop which will determine the latency of \texttt{task} loop in both the original and transformed codes. This feature is helpful for us to analyze the performance bottleneck and identify the key tuning parameter by running HLS once at each iteration instead of evaluating the effect of all $K$ parameters.

By exploiting the cycle breakdown, we can resolve the issues mentioned above by developing a bottleneck analyzer. We first build a map from the loop or function statements in the user input code to design parameters so that we know which parameters should be focused on for a particular statement. To identify the critical path and type, we start with the kernel top function statement and build hierarchy paths of the design by traversing the Merlin report using depth-first search (DFS). More specifically, for each hierarchy level, we first check to see if the current statement has child loop statements and sort them by their latency. Then, we traverse each of the child loops and repeat this process. In case of a function call statement, we dive into the function implementation to further check its child statements for building the hierarchy paths. Finally, we return a list of paths in order.
Note that since we sort all loop statements according to their latency by checking the Merlin report, the hierarchy paths we created will also be sorted by their latency. 

Subsequently, for each statement, we check the Merlin report again to determine whether its performance bottleneck is memory transfer or computation. The Merlin Compiler obtains this information by analyzing the transformed kernel code along with the HLS report. A cycle is considered to be a memory transfer cycle if it is consumed by communicating to global memory. As a result, we can not only figure out the performance bottleneck for each design point, but also identify a small set of effective design parameters to focus on. Therefore, we are able to significantly improve the efficiency of our searching algorithm.

When we obtain an ordered list of critical hierarchy paths from the bottleneck analyzer, we start from the innermost loop statement (because of the DSF traversal) of the most critical entry and identify its corresponding parameters as candidate parameters to explore, if they are not already tuned. 
Based on the bottleneck type, provided by the bottleneck analysis, (i.e., memory transfer or computation), we pick a subset of the parameters mapped to that statement to work on. For example, we may have design parameters of \texttt{PARALLEL} and \texttt{TILING} at the same loop level. When the bottleneck type of the loop is memory transfer, we focus on the \texttt{TILING} parameter for the loop; otherwise, we focus on \texttt{PARALLEL} parameter. In other words, we reduce the number of candidate design parameters not only by the bottleneck statement but also by the bottleneck type.

We define each design point as a data structure containing the following information:
\vspace{0.05in}
\begin{lstlisting}[numbers=none, language=python]
curr_point = (*@\textcolor{blue}{DesignPoint}@*)(configuration, tuned, result, quality, children)
\end{lstlisting}
where \textit{configuration} contains the value of all the parameters and \textit{tuned} lists the parameters which the algorithm has explored for the current point. \textit{quality} stores the quality of design measured by finite difference value and \textit{result} includes all the related information gathered from the HLS tool including the resource utilization and the cycle count. Finally, each design point stores a stack of the configurations for its unexplored children where each child is generated by advancing one of the parameters by one step. The children are pushed to the stack in the order of their importance (from least to most important) as computed by the bottleneck analyzer so that by popping the stack, we get to work with the child who has changed the parameter with the most promising impact.

We define level \textit{n} as a point where we have fixed the value of \textit{n} parameters, so the maximum level in our algorithm is equal to the total number of parameters. For each level, we define a heap of the pending design points that can be further explored and push the design points by their \textit{quality} into the heap. Since new design points are sorted by their quality values when they were pushed into the heap, the design point with a better quality value will be chosen for tuning more of its parameters prior to other points. As mentioned above, the next point to be explored is chosen by popping the stack of the unexplored children of this design point so that at each step, we get to evaluate the most promising design point.

\begin{algorithm}
  \caption{$\framework$ Explorer: Bottleneck-guided Coordinate Optimizer}
  \label{alg:bottleneck}
  \begin{algorithmic}[1]
  
  \REQUIRE A C program $\mathcal{P}$ and a set of design space partitions $\mathbb{P}$.
  \ENSURE A design configuration $\theta$ with the best QoR.

  \STATE $top\_func \leftarrow GetTopFunction(\mathcal{P})$
  \FORALL{$P \in \mathbb{P}$}
    \STATE $best\_ cfg = cfg \leftarrow GetDefaultPoint(P)$
    \STATE $report, hier \leftarrow Evaluate(cfg)$
    \STATE $parameter\_order \leftarrow BottleneckAnalysis(report, hier, top\_func, \varnothing)$
    \STATE $children \leftarrow GetChildren(cfg, parameter\_order)$
    \STATE $LevelHeap \leftarrow \varnothing$
    \STATE $LevelHeap.append(\varnothing)$
    \STATE $LevelHeap[0].push(DesignPoint(cfg, \varnothing, 0, report.result, children))$
    \WHILE{$LevelHeap \notin \varnothing \; \textbf{and} \; elapsed\_time < DSE\_TIMEOUT $}
        \STATE $curr\_level = GetLastLevel(LevelHeap)$
        \STATE $curr\_point \leftarrow LevelHeap[curr\_level].peek()$
        \STATE $tuned\_parameters = curr\_point.tuned$
        \STATE $candidate\_cfg, focused\_parameter \leftarrow curr\_point.children.pop()$
        \FORALL{$option \in focused\_parameter$}
        \STATE $new\_cfg \leftarrow Manipulate(candidate\_cfg, focused\_parameter, option)$
        \STATE $new\_tuned \leftarrow tuned\_parameters + [(focused\_parameter, option)]$
        \STATE $report, hier \leftarrow Evaluate(new\_cfg)$
        \STATE $quality \leftarrow CalQuality(report.result, \myquote{FiniteDifference})$
        \STATE $best\_cfg \leftarrow UpdateBest(new\_cfg, quality)$
        \STATE $parameter\_order \leftarrow BottleneckAnalysis(report, hier, top\_func, new\_tuned)$
        \IF{$len(parameter\_order)$ > 0}
            \STATE $children \leftarrow GetChildren(new\_cfg, parameter\_order)$
            \STATE $new\_point \leftarrow DesignPoint(new\_cfg, new\_tuned, quality, report.result, children)$
            \STATE $LevelHeap[curr\_level+1].push(new\_point)$
        \ENDIF
        \ENDFOR
        
        \IF{$LevelHeap[curr\_level].peek().NumChildren == 0$}
            \STATE $LevelHeap[curr\_level].pop()$
        \ENDIF
    \ENDWHILE
  \ENDFOR
  \RETURN $best\_cfg$
  \end{algorithmic}
\end{algorithm}
Algorithm~\ref{alg:bottleneck} presents our exploring strategy. As we will explain in Section~\ref{sec:gradient_partition}, we partition the design space to alleviate the local optimum problem. For each partition, we first get its default point and initialize the heap of the first level (lines 3 to 9). Then, at each iteration of the algorithm, $\framework$ gets the heap with the highest level, peeks the first node of the heap, and pops its stack of unexplored children to get the new candidate (lines 11 to 14). Next, each option of the new focused parameter will be evaluated and the result will be passed to the bottleneck analyzer to generate a new set of focused parameters for making new children (lines 16 to 21). Since the number of fixed parameters is increased by one, it will be pushed to the heap of the next level if there is still a parameter left that has not been tuned yet (lines 22 to 26).
When the stack of unexplored children of the current design point is empty, it will be popped out of heap (lines 28 to 30). The algorithm continues either until all the heaps are empty or when the DSE has reached a runtime threshold (Line 10).

As an example, when {$\framework$} optimizes Code~\ref{code:cnn_hls_tranformed}, it will see that the \textit{convolution} part of the code takes 85.2\% of the overall cycle counts. Since that section of the code is a series of nested loops, the parameters of the inner-most loop will take the top of the list produced by the bottleneck analyzer. We explain in Section~\ref{sec:gradient_ds} that we do not consider loops with trip count of less than or equal to 16 in our DSE since the HLS tool can automatically optimize these loops well. As a result, the \texttt{w} loop in Line 15 would be the inner-most loop with parameters which the Merlin report tells us it is a computation-bound loop. As we describe Section~\ref{sec:param-order}, {$\framework$} first tries to apply \texttt{fg PIPELINE} which would be a successful attempt. In the next iteration, the last level heap will contain the design point that was just optimized and since the \textit{convolution} part is still the bottleneck, {$\framework$} would try parallelizing the \texttt{w} loop and will choose \texttt{factor=4} since it achieves the highest \textit{quality} value. Although \texttt{factor=8} can reduce the cycle count by 11\%, it increases the overall area (Eq.\ref{eq:util}) by 63\% which results in a worse quality; therefore, {$\framework$} picks \texttt{factor=4} to make room for further improvement. By adopting Algorithm~\ref{alg:bottleneck}, {$\framework$} can improve the performance by $218\times$ very quickly, only after 2 iterations of the algorithm.

\subsection{Parameter Ordering} \label{sec:param-order}
It often happens that each bottleneck type has more than one applicable design parameter. In these situations, we sort the parameters by a pre-defined priority. For example, if the bottleneck of a loop statement is determined to be its computation, one can apply \texttt{fg} or \texttt{cg} pipelining/parallelization, in general. 
In this case, we treat the \texttt{PIPELINE} pragma as two different parameters based on its mode and choose the order of applying the pragmas to be \texttt{fg PIPELINE}, {PARALLEL}, and {cg PIPELINE} which is a heuristic approach to improve the performance by utilizing more fine-grained parallelization units since the HLS tool handles such optimizations better. Here, measuring the quality of design points with the finite difference value (Eq.~\ref{eq:fd}) helps $\framework$ not to over-utilize the FPGA. For a configuration, when the gain of the achieved speedup is not comparable to the loss of available resources, the quality of design decreases; hence, $\framework$ will not tune that parameter and the resources are left for applying a design parameter with higher impact. 

\begin{table}[!tbh]
\centering
\caption{Performance and Area Compared to The Base Design When Parameters of Line 15 in Code~\ref{code:cnn_hls_tranformed} Change. TIMEOUT is set to 60 minutes. The results suggest that applying fine-grained optimization first lets the HLS tool synthesize the design easier.}
\label{tbl:resource_cnn}
\begin{tabularx}{\linewidth}{|c|c|Y|Y|Y|Y|Y|}
\hline
Optimization   & Status & Perf & BRAM &LUT  & DSP  & FF   \\ \hline\hline
Pi-\texttt{fg}         & PASS (24 min) & 175$\times$ & +7\% & +23\% & +24\%  & +15\%  \\ \hline
PF=4         & TIMEOUT & - & - & - & -  & -  \\ \hline
Pi-\texttt{fg} + PF=4       & PASS (28 min) & 218$\times$ & +17\%  & +44\% & +33\% & +25\% \\ \hline
\end{tabularx}
Pi: Pipeline, PF: Parallel Factor, fg: fine-grained
\end{table}

Moreover, as mentioned in Challenge 3 of Section~\ref{sec:intro}, the order of applying the pragmas is crucial in order to get to the best design. Our experiments show that evaluating the fine-grained optimizations first helps $\framework$ reach the best design point in fewer iterations. This is mainly because HLS tools schedule fine-grained optimizations better than the coarse-grained ones. Table~\ref{tbl:resource_cnn} shows how the performance and resource utilization change when \texttt{fg PIPELINE} and {PARALLEL} pragmas are applied to line 15 in Code~\ref{code:cnn_hls_tranformed} compared to the base design where all the pragmas are off. The time limit to run the HLS tool is set to 60 minutes. The results suggest that in order to get to the optimal configuration for this loop, we must first apply the fine-grained pipelining. This way, the HLS tool can better schedule the loop when parallelization is further applied and its synthesis will finish in 28 minutes. However, if we first apply the other pragma which results in a coarse-grained parallelization, the synthesis will be timed out and {$\framework$} does not tune this pragma at this stage.

Note that we do not prune the other design parameters. We just change the order of the parameters to be explored as these rules can not be generalized to all cases due to the unpredictability of the HLS tools. If the bottleneck of a design point is memory transfer, $\framework$ prioritizes \texttt{cg PIPELINE} over \texttt{TILING} pragma. The Merlin Compiler, by default, caches the data and the former will further overlap the communication time with computation by applying double buffering; however, the latter, can be used to change the size of the cached data.

\subsection{Efficient Design Space Representation} \label{sec:gradient_ds}
To further facilitate the bottleneck-based optimizer, we seek to reduce the ineffective parameters. Intuitively, we can build a grid design space from the Merlin pragmas by treating each pragma as a tuning parameter and search for the best combination. However, many points in this grid space may be infeasible. For example, if we have determined to perform coarse-grained pipelining at the outermost loop of a loop nest, the Merlin Compiler will apply double-buffering on the loop. In this case, the physical meaning of double-buffering at the outermost loop is to transfer a batch of data from DRAM to BRAM, which cannot be further parallelized. As a result, pipeline and parallel pragmas are mutually exclusive in a loop nest. We propose an efficient approach to create a design space that preserves the grid design space but invalidates infeasible combinations.

\begin{figure}[!tbh]
	\centering
	\includegraphics[scale=0.8]{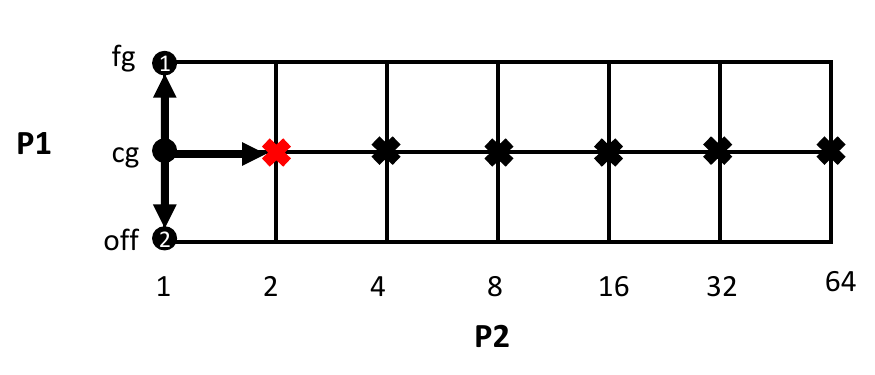} 
	\caption{Proposed Design Space Representation and Its Impact on DSE. P1 and P2 denote the \texttt{PIPELINE} and \texttt{PARALLEL} pragmas, respectively}
	\label{fig:bottleneck_ds}
\end{figure}

Fig.~\ref{fig:bottleneck_ds} illustrates the goal of an efficient design space representation. 
In this example, we attempt to explore the best parameter with the best option for loop \texttt{j} of Code~\ref{code:cnn_hls_tranformed} with pragma P1 and P2 denoting the \texttt{PIPELINE} and \texttt{PARALLEL} pragmas, respectively. Pragma $P1$ and $P2$ are exclusive when $P1$ is used in \texttt{cg} mode; therefore, only one of them should be inserted at a time. 
A good design space representation must preserve the grid design space but invalidate infeasible points. An example of such representation is presented in Fig.~\ref{fig:bottleneck_ds}. Assume that we are at the configuration $(P1, P2)=(\texttt{cg}, 1)$, we only have two candidates to explore in the next step because the configuration $(P1, P2)=(\texttt{cg}, 2)$ is invalid. This representation is exploration friendly and, it is easy to enforce rules on the infeasible points.

To represent a grid design space with invalid points, we introduce a \textit{Python} list comprehension syntax to {$\framework$}. The \textit{Python} list comprehension is a concise approach for creating lists with conditions. It has the following syntax:

\vspace{0.05in}
\begin{lstlisting}[language=python,numbers=none]
list_name = [expression for item in list if condition]
\end{lstlisting}

\noindent Formally, we define the design space representation for Merlin pragmas with list comprehensions as follows:
\vspace{0.05in}
\begin{lstlisting}[numbers=none]
#pragma ACCEL <pragma-type> <attribute-key>=auto{
  options: parameter_name=list-comprehension-expression;
  default: default-value }
\end{lstlisting}

For our example, the design space can be represented using list comprehensions as follows:
\vspace{0.05in}
\begin{lstlisting}[language=C]
// Skip the rest due to page limit
#pragma ACCEL PIPELINE auto{
  options: P1 = [x for x in [off, cg, fg]];
  default: 'off' }
#pragma ACCEL PARALLEL factor=auto{
  options: P2 = [x for x in [1, 2, 4, 8, 16, 32, 64] if P1!=cg];
  default: 1 }
for (int j = 0; j < NumIn; ++i) {
// Skip the rest due to page limit
\end{lstlisting}

\noindent where line 6 indicates that the two pragmas are exclusive. In other words, when we set $P1=\texttt{cg}$, the available option for $P2$ is only the default value, which is $1$ in this case. Note that the default value of each pragma turns it off.

There are three main advantages to adopting list comprehension-based design space representations. First, we are able to represent a design space with exclusive rules to greatly reduce its size. Second, the \textit{Python} list comprehension is general. It provides a friendly and comprehensive interface for higher layers such as polyhedral analysis~\cite{polyframework} and domain-specific languages to generate an effective design space in the future. Third, the syntax of this representation is \textit{Python} compatible. This means we can leverage the \textit{Python} interpreter to evaluate the design space and improve overall stability of the DSE framework. 

The Design Space Generator, depicted in Fig.~\ref{fig:framework}, adapts the Rose Compiler~\cite{rose} to analyze the kernel AST and extract the required information for running the DSE such as the loops in the design, their trip-count, and available bit-width. Artisan~\cite{vandebon2020artisan} employs a similar approach for analyzing the code. However, it only considers unroll pragma in code instrumentation. Our approach, on the other hand, considers a wider set of pragmas as mentioned in Table~\ref{tbl:merlin_pragmas} and exploits the following rules to prune the design space: 
\begin{itemize}
\item Ignore the fine-grained loops with trip count (TC) of less than or equal to 16 as the HLS tool can schedule these loops well.
\item Tiling factors (TF) should be integer divisors of their loop TC.
\item The allowed parallel factors (PF) for a loop are all sub-divisors of the loop TC up to $min(128, TC)$ plus the TC itself. PF of larger than 128 causes the HLS tool to run for a long time and it usually does not result in a good performance.
\item For each loop, we should have $TF * PF \le TC$.
\item When \texttt{fg PIPELINE} is applied on a loop, no other pragma is allowed for the inner loops since this parameter want to unroll all the inner loops completely.
\item A parallel pragma is invalid for a loop nest when \texttt{cg PIPELINE} is applied on that loop.
\item A tiling pragma is added only to the loops with an inner loop.
\end{itemize}

\subsection{Design Space Partitioning} \label{sec:gradient_partition}
Unfortunately, the third inefficiency of the approaches reviewed in Section~\ref{sec:learn} also exists in our bottleneck-guided optimizer. We still cannot identify whether the current option of a parameter is locally or globally optimum. The most promising solution is breaking the dependency between options and searching a set of them in parallel. Although we need to evaluate multiple design points at every iteration, each design point will provide the maximum information for improving the performance because we always evaluate the parameters that have the largest impact on the performance bottleneck.

By partitioning the design space based on the likely distribution of locally optimal points and exploring each partition independently, we solve the local optimum issue caused by the non-smooth performance trend (Challenge 2 in Section~\ref{sec:intro}) since each partition starts exploring from a different point. Intuitively, we could partition the design space according to the range of values of every parameter in a design, but it may generate thousands of partitions and result in a long exploration time. Instead, we partition the design space based on the pipeline mode, as \texttt{fg PIPELINE} unrolls all sub-loops while the \texttt{cg PIPELINE} exploits double buffers to implement coarse-grained pipelining. These two modes apparently have the most significant different influence on the generated architecture and are expected to have non-related performance and resource utilization. According to the pipeline modes in each loop, we use the tree partition and generate $2^{m}$ partitions from a design space with $m$ non-innermost loops.

Supposing we use $t$ working threads to perform, at most, $h$ hours DSE for $2^{m}$ design space partitions, we need $\frac{2^m}{t} \times h$ hours to finish the entire process. On the other hand, some partitions that are based on an insignificant pipeline pragma may have a similar performance, so it is more efficient to only explore one of them. As a result, we profile each partition by running HLS with minimized parameter values to obtain the minimum area and performance and use K-means clustering with performance and area as features to identify $t$ representative partitions among all $2^{m}$ partitions. 

\section{Evaluation} \label{sec:eval}

\subsection{Experimental Setup}
Our evaluation is performed on Amazon Elastic Compute Cloud (EC2)~\cite{amazon-f1}. We use \texttt{r4.4xlarge} instance with 16~cores and 122~GiB memory to perform the DSE and generate accelerator bit-streams. The generated FPGA accelerators are evaluated on an F1 instance (\texttt{f1.2xlarge}) with Xilinx Virtex UltraScale+\textsuperscript{TM} VU9P FPGA. In addition, we choose the commonly-used MachSuite~\cite{machsuite} benchmark suite and the FPGA-friendly Rodinia~\cite{che2009rodinia} benchmark, along with one convolution layer of Alexnet~\cite{alexnet} as our first benchmark. For several common kernels, MachSuite provides C implementation that is programmed without the consideration of FPGA acceleration, which makes it a natural fit for demonstrating our approach. We evaluate the effect of our optimizations and compare the designs generated by our tool to the state-of-the-art works using this benchmark. Furthermore, to the best of our knowledge, we are the first ones to evaluate the performance of our tool on vision kernels of Xilinx Vitits libraries~\cite{vitis} that are optimized for Xilinx FPGAs, based on the OpenCV library~\cite{opencv_library}.

\subsection{Evaluation of Optimization Techniques}
\begin{figure*}[!htb]
	\centering
	\includegraphics[width=\linewidth]{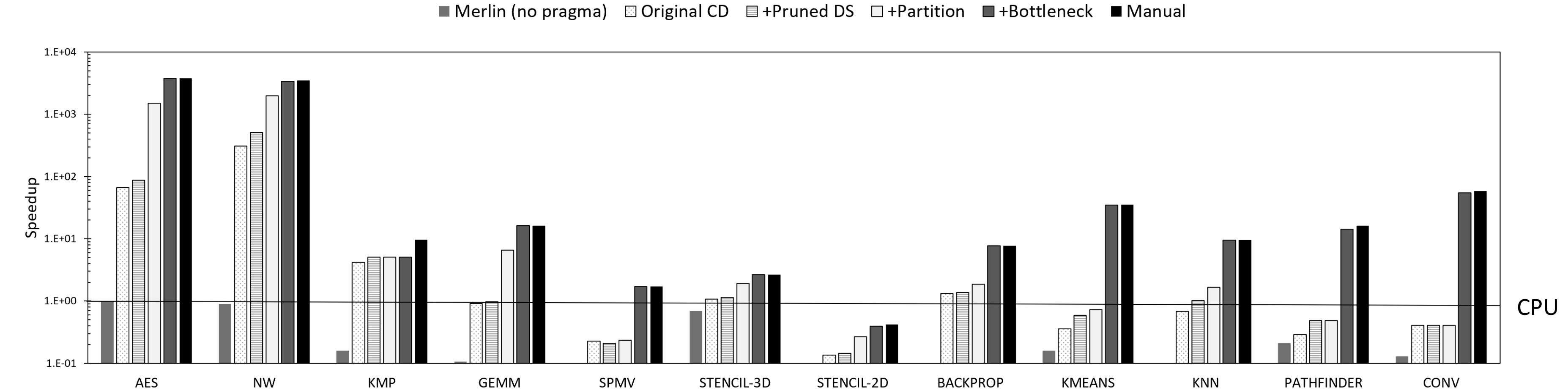} 
	\caption{Speedup of the Merlin Compiler without any Pragmas, Proposed Approach with Different Optimizations, and the Manual Design over an Intel Xeon CPU Core}
	\label{fig:exp_approach}
\end{figure*}

We first measure the performance of the Merlin Compiler without any pragmas and without the help of AutoDSE to get the impact of its default optimizations. The $1^{st}$ bar of each case in Fig.~\ref{fig:exp_approach} depicts the speedup gained by the Merlin Compiler with respect to CPU. Then, we evaluate the original coordinate descent (CD) method described in Section~\ref{sec:learn} and the proposed optimization strategies explained in sections~\ref{sec:gradient_ds} and~\ref{sec:gradient_partition}. The $2^{nd}$ to $4^{th}$ bars in Fig.~\ref{fig:exp_approach} show the speedup gained after tuning the pragmas by each of these optimizations. Note that the chart is in logarithmic scale. We can see that the default optimizations of the Merlin Compiler are not enough and after applying the candidate pragmas generated by the Original CD, we get $13.52\times$ speedup, on the geometric mean. Moreover, each of the proposed strategies benefits at least one case in our benchmark and together further bring a $2.47\times$ speedup. The list-based design space representation keeps the search space smooth by invalidating infeasible combinations. As a result, we can investigate more design points in a fixed amount of time. This helps \texttt{AES}, \texttt{NW}, \texttt{KMP}, \texttt{PATHFINDER}, \texttt{KMEANS}, and \texttt{KNN}.
Design space partition benefits the designs with many loop nests in which the coordinate process is easily trapped by the local optimum when changing pipeline modes---such as \texttt{AES}, \texttt{GEMM}, \texttt{NW}, \texttt{STENCIL-2D}, and \texttt{STENCIL-3D}.

The $5^{th}$ bar shows the speedup of $\framework$ when the bottleneck-guided coordinate optimizer detailed in Section~\ref{sec:bottleneck_algo} is adapted along with the parameter ordering explained in Section~\ref{sec:param-order}, design space representation introduced in Section~\ref{sec:gradient_ds}, and design space partitioning described in Section~\ref{sec:gradient_partition}. With this setup, $\framework$ further improves the result by 5.5$\times$ on the geometric mean bringing the overall speedup compared to when no pragmas are applied to $182.92\times$. As a result, $\framework$ is able to achieve a speedup of \speedupcpu$\times$ over CPU and get to 0.93$\times$ performance of the manual designs while running only for 1.1 hours on the geometric mean. The manual designs, depicted by the $6^{th}$ bar, are optimized by applying the Merlin pragmas \textit{manually} without changing the source programs.

\begin{figure}[!tbh]
	\centering
	\includegraphics[scale=0.3]{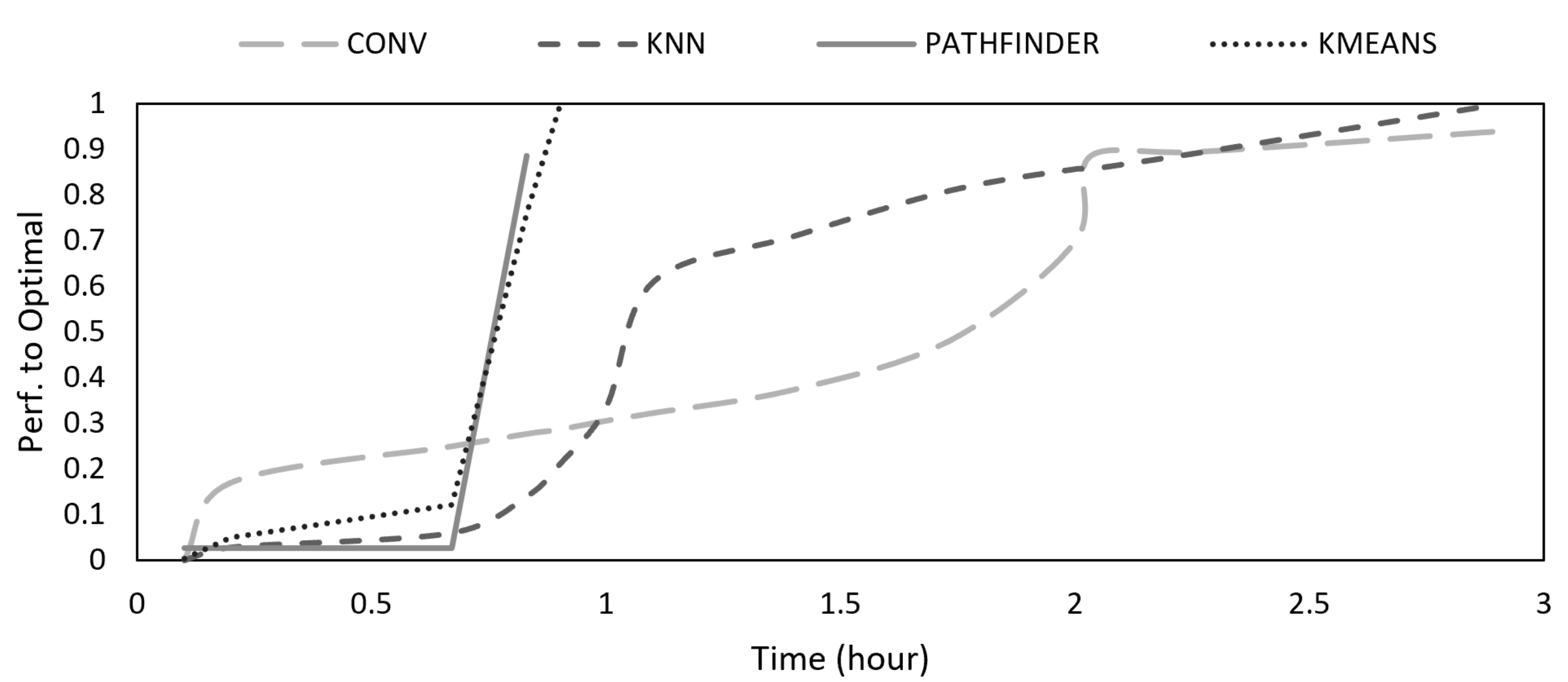} 
	\caption{Speedup Over the Manual Design Using {\framework} for the 4 Cases that the Bottleneck-guided Optimizer had Significant Impact}
	\label{fig:exp_trend_hs}
\end{figure}

Fig.~\ref{fig:exp_trend_hs} depicts the $\framework$ process for four cases where the bottleneck-guided optimizer showed significant improvement in the performance. This shows that our approach can rapidly achieve a high performance design. $\framework$ does not exactly match the performance of manual designs for all of the cases because the HLS report may not reflect the accurate computation cycles when the kernels contain many unbounded loops or while-loops, which in turn affects the Merlin report. In order to get the importance of the parameters, the bottleneck analyzer (explained in Section~\ref{sec:bottleneck_algo}) needs to receive the accurate cycle estimation of the design. In the absence of the true cycle breakdown, it cannot determine the high-impact design parameters. Therefore, our search algorithm may focus on unimportant parameters. 

\subsection{Comparison with Other DSE Approaches} \label{sec:exp-other-dse}
We further evaluate the overall performance of generated accelerator designs by {$\framework$} compared to the previous state-of-the-art works including S2FA~\cite{s2fa}, lattice-traversing DSE~\cite{ferretti2018lattice}, and Gaussian process-based Bayesian optimization~\cite{date21} in Table~\ref{tbl:comparison}. The numbers show the speedup of the design found by {$\framework$} compared to the design that their framework found after running the tools for the same allotted time. Note that the performance of the other works are not reported by the authors for all of the kernels we are testing. According to Table~\ref{tbl:comparison}, by utilizing the bottleneck approach, we can outperform S2FA, lattice-traversing DSE, and Gaussian process-based Bayesian optimization by 3.6$\times$, 4.3$\times$, 17.9$\times$ respectively, on the geometric mean. 

\begin{table}[!tbh]
\centering
\caption{Speedup of Our Approach Compared to S2FA~\cite{s2fa}, Lattice-traversing DSE~\cite{ferretti2018lattice}, and Gaussian process-based Bayesian Optimization~\cite{date21}}
\label{tbl:comparison}
\begin{tabular}{|c||c|c|c|c|c|c||c|}
\hline
\rule{0pt}{4ex}  Approach   & AES & NW & GEMM  & KMP  & SPMV & STENCIL-3D  & \textbf{GEO-Mean} \\ \hline\hline
Lattice (ICCD'18)~\cite{ferretti2018lattice}        & 1.63 & 6.32 & 7.39 & -  & - & - & \textbf{4.3} \\ \hline
S2FA (DAC'18)~\cite{s2fa}        & 512.86 & 1  & 1.52 & 1.74 & 1 & 1.26 & \textbf{3.6} \\ \hline
Bayesian (DATE'21)~\cite{date21}         & - & - & 100.17 & -  & 2.07 & 27.75 & \textbf{17.9} \\ \hline
\end{tabular}
\end{table}


As we discussed in Section~\ref{sec:learn}, the deficiency of S2FA stems from how hard it is for the problem-independent learning algorithm to find the key parameters. Lattice-traversing DSE needs an initial sampling step to learn the design space. This takes a long time for our benchmark due to the size of the design space even though the authors only consider unrolling the loops and function inlining. This constraint makes it hard for the tool to start the exploration process before the time limit for DSE is met. The Gaussian process-based Bayesian optimization also has to spend some time to sample the design space and build an initial surrogate model. However, {$\framework$} can learn the high-impact directives by exploring the performance breakdown and thus, is able to find a high-performance design in a few iterations.

Moreover, adopting the Merlin Compiler as the backend gives further advantage to {$\framework$} compared to other DSE tools. This allows the tool to exploit the automatic code transformations for applying the common optimization techniques such as memory burst, memory coalescing, and double buffering; and focus only on high-level hardware changes. Nonetheless, the performance comparison with S2FA demonstrates that adopting the Merlin Compiler is not enough and we still need to explore the design space more efficiently.

\subsection{Comparison with Expert-level Manual HLS Designs} \label{sec:exp_vitis}

To further evaluate the performance of $\framework$, we use 33 vision kernels from Xilinx Vitis Library~\cite{vitis}. These kernels utilize 14 optimization pragmas, on average (by the geometric mean), which include \texttt{UNROLL}, \texttt{PIPELINE}, \texttt{ARRAY\_PARTITION}, \texttt{DEPENDENCE}, \texttt{LOOP\_FLATTEN}, \texttt{INLINE}, \texttt{DATAFLOW}, and \texttt{STREAM}. For each kernel, we remove all the optimization pragmas except for \texttt{DATAFLOW} and \texttt{STREAM}. The removed pragmas, which are of the first six types mentioned above, are used 13.47 times on average (out of 14). As a result, we require less than one optimization pragma per kernel, on the geometric mean. The only optimization pragmas kept are \texttt{DATAFLOW} and \texttt{STREAM} pragmas. This is because our search space is built on top of the Merlin Compiler and we do not search for the \texttt{DATAFLOW} and \texttt{STREAM} pragmas as these pragmas are not among the Merlin-specified pragmas. In the future, we will expand our search engine to HLS pragmas that are not included in Merlin. Furthermore, the \texttt{INTERFACE} and \texttt{LOOP\_TRIPCOUNT} pragmas are also kept which are not among the HLS optimization pragmas. They are rather used to specify the connection to AXI bus and the range of the trip count of the loop, respectively.

\begin{table}[!tbh]
\centering
\caption{Average (Geometric Mean) Speedup of the Vitis tool, the Merlin Compiler, and {$\framework$} over the Manually Optimized Kernels from Xilinx Vitis Libraries. The manual designs are the original kernels from the library. The performance of those designs are compared to when the optimization pragmas we search for (\texttt{UNROLL}, \texttt{PIPELINE}, \texttt{ARRAY\_PARTITION}, \texttt{DEPENDENCE}, \texttt{LOOP\_FLATTEN}, and \texttt{INLINE}) are removed and the code is passed to three different tools.}
\label{tbl:exp-vitis}
\begin{tabular}{|c||c||c|c|c|}
\hline
\backslashbox{Comparison Scenario}{Compared to}  & {\makecell{Vitis \\ (Manually Optimized)}} & {\makecell{Vitis \\ (Default)}}   & Merlin Compiler & {$\framework$}   \\ \hline\hline
\makecell{Speedup over the Vitis Library with \\ (Original) Manually Inserted Pragmas} & $1\times$ & \makecell{$0.12\times$} & \makecell{$0.38\times$} & \makecell{$1.04\times$} \\ \hline
\makecell{Performance Improvement over the \\ Vitis Tool with Default Settings} & $8.69\times$ & $1\times$ & $3.29\times$ & $9.04\times$ \\ \hline\hline
\#pragmas Listed in the Table's Caption & 13.47 & 0 & 0 & 0 \\ \hline
Total Pragma Reduction & $1\times$ & \pragmareductionvitis & \pragmareductionvitis & \pragmareductionvitis \\ \hline
\end{tabular}
\end{table}

To better understand the effect of our optimizer, we tested the performance of the Vitis tool and the Merlin Compiler on the input to {$\framework$} (which does not include the optimization pragmas mentioned above). The performance comparisons are summarized in Table~\ref{tbl:exp-vitis}. As the results show, while the Merlin Compiler can get to a speedup of $3.29\times$ compared to the Vitis tool, it still needs the help of {$\framework$} to get to the manually optimized kernels in the library. In fact, {$\framework$} could achieve a further speedup of 2.74$\times$ by automatically inserting 3.2 Merlin pragmas per kernel, on the geometric mean. As a result, it could improve the performance of the Vitis tool by $9.04\times$ and $1.04\times$ when the code with reduced set of pragmas and the manual code, respectively, are passed to it.

Fig.~\ref{fig:exp_vitis} in Appendix~\ref{sec:a2_vitis} depicts the performance comparison of the design points $\framework$ generated with respect to Xilinx results along with the number of pragmas that we removed in detail. The results show that {$\framework$} is able to achieve to a same or better performance yet with {\pragmareductionvitis} reduction of their optimization pragmas in 0.3 hours, on the geometric mean; therefore, proving the effectiveness of our bottleneck-based approach and the fact that it can mimic the method an expert would take. For the cases that $\framework$ does not exactly match the performance of Vitis, {$\framework$} still finds the best combination of the pragmas. The inequality lies in the different II that Merlin has achieved. For example, the \texttt{histEqualize}, \texttt{histogram}, and \texttt{otsuthreshold} kernels all have a loop that requires the II to be set to 2 when \texttt{PIPELINE} pragma is used. Otherwise, Vivado HLS achieves an II=3. However, it is not possible to change the II using the Merlin Compiler. On the other hand, $\framework$ is able to outperform the performance of \texttt{customConv} and \texttt{reduce} kernels significantly by better detecting the choices and locations for pipelining and parallelization.
\vspace{-0.15cm}
\section{Conclusion and Future Work} \label{sec:conclusion}
In this paper, we made our first, yet very important, step of lowering the bar of accelerating programs using FPGA for general software programmers to make FPGA universally accessible.
We analyzed the difficulty of exploring HLS design space. To address challenges 2 to 4 mentioned in Section~\ref{sec:intro}, we treat the HLS tool as a black-box. We use the synthesis results to estimate the QoR rather than the placement and routing ($P\&R$) results, because $P\&R$ is too time-consuming to explore sufficient design points in a reasonable time budget. According to our observation and analysis, we propose a bottleneck-guided coordinate optimizer and develop a push-button framework, {$\framework$}, based on that to systematically approach a better solution. By exploring the solution space efficiently, we address challenges 1 and 5. We propose a heuristic for ordering the parameters that can further help challenges 3 and 5. To eliminate meaningless design points, we incorporate a list comprehension-based design space representation and prune 24.65$\times$ ineffective configurations on average, while keeping the design space smooth; hence, further alleviating Challenge 1. Additionally, we employ a partitioning strategy to address the local optimum problem mentioned in Challenge 2. We show that {$\framework$} can outperform general hyper-heuristics used in the literature by focusing on high-impact design parameters first. The experimental results suggest that $\framework$ lets anyone with a decent knowledge of programming try customized computing with minimum effort. 

{$\framework$} is built with the assumption that we can get the performance breakdown of the program from the HLS tool. We expect all HLS tools will provide performance breakdown at some point, as it is important for manual performance optimization (such as the need for Intel VTune Profiler~\cite{vtune} in the case of CPU performance optimization). Xilinx HLS is already providing such information that the Merlin Compiler leverages and Intel OpenCL~\cite{intel-sdk} is planning to add this feature. It is likely that other HLS tools~\cite{cadence,catapult-hls,nec} will add such information as well in the near future. Hence, we believe, it is reasonable for {$\framework$} to take advantage of such information to mimic the human performance optimization process to perform bottleneck-driven DSE.

In the future, we plan to include more transformations (design space parameters) for optimizing data access and reuse patterns. We will also extend {$\framework$} to estimate the QoR based on the $P\&R$ results by developing a machine learning model to predict them from the synthesis results as in~\cite{dai2018fast}.

\begin{acks}
The authors would like to thank Dr. Peichen Pan for his invaluable support with the Merlin Compiler and Dr. Lorenzo Ferretti and Qi Sun for helping with the comparison to their work. This work is supported by the ICN-WEN award jointly funded by the NSF (CNS-1719403) and Intel (34627365), the CAPA award also jointly funded by NSF (CCF-1723773) and Intel (36888881), and CDSC industrial partners\footnote{https://cdsc.ucla.edu/partners/}.
\end{acks}

\bibliographystyle{ACM-Reference-Format}
\bibliography{main}


\begin{thebibliography}{63}


\ifx \showCODEN    \undefined \def \showCODEN     #1{\unskip}     \fi
\ifx \showDOI      \undefined \def \showDOI       #1{#1}\fi
\ifx \showISBNx    \undefined \def \showISBNx     #1{\unskip}     \fi
\ifx \showISBNxiii \undefined \def \showISBNxiii  #1{\unskip}     \fi
\ifx \showISSN     \undefined \def \showISSN      #1{\unskip}     \fi
\ifx \showLCCN     \undefined \def \showLCCN      #1{\unskip}     \fi
\ifx \shownote     \undefined \def \shownote      #1{#1}          \fi
\ifx \showarticletitle \undefined \def \showarticletitle #1{#1}   \fi
\ifx \showURL      \undefined \def \showURL       {\relax}        \fi
\providecommand\bibfield[2]{#2}
\providecommand\bibinfo[2]{#2}
\providecommand\natexlab[1]{#1}
\providecommand\showeprint[2][]{arXiv:#2}

\bibitem[\protect\citeauthoryear{??}{ros}{[n.d.]}]%
        {rose}
 \bibinfo{year}{[n.d.]}\natexlab{}.
\newblock \showarticletitle{{Rose Compiler Infrastructure}}.
\newblock
\newblock
\shownote{http://rosecompiler.org/.}


\bibitem[\protect\citeauthoryear{{Amazon EC2 F1 Instance}}{{Amazon EC2 F1
  Instance}}{[n.d.]}]%
        {amazon-f1}
\bibfield{author}{\bibinfo{person}{{Amazon EC2 F1 Instance}}.}
  \bibinfo{year}{[n.d.]}\natexlab{}.
\newblock \bibinfo{title}{\url{https://aws.amazon.com/ec2/instance-types/f1/}}.
\newblock
\newblock


\bibitem[\protect\citeauthoryear{Andrade, George, Karras, Novo, Pratas, Sousa,
  Ienne, Falcao, and Silva}{Andrade et~al\mbox{.}}{2017}]%
        {andrade2017design}
\bibfield{author}{\bibinfo{person}{Joao Andrade}, \bibinfo{person}{Nithin
  George}, \bibinfo{person}{Kimon Karras}, \bibinfo{person}{David Novo},
  \bibinfo{person}{Frederico Pratas}, \bibinfo{person}{Leonel Sousa},
  \bibinfo{person}{Paolo Ienne}, \bibinfo{person}{Gabriel Falcao}, {and}
  \bibinfo{person}{Vitor Silva}.} \bibinfo{year}{2017}\natexlab{}.
\newblock \showarticletitle{Design space exploration of LDPC decoders using
  high-level synthesis}.
\newblock \bibinfo{journal}{\emph{IEEE Access}}  \bibinfo{volume}{5}
  (\bibinfo{year}{2017}), \bibinfo{pages}{14600--14615}.
\newblock


\bibitem[\protect\citeauthoryear{Ansel, Kamil, Veeramachaneni, Ragan-Kelley,
  Bosboom, O'Reilly, and Amarasinghe}{Ansel et~al\mbox{.}}{2014}]%
        {opentuner}
\bibfield{author}{\bibinfo{person}{Jason Ansel}, \bibinfo{person}{Shoaib
  Kamil}, \bibinfo{person}{Kalyan Veeramachaneni}, \bibinfo{person}{Jonathan
  Ragan-Kelley}, \bibinfo{person}{Jeffrey Bosboom}, \bibinfo{person}{Una-May
  O'Reilly}, {and} \bibinfo{person}{Saman Amarasinghe}.}
  \bibinfo{year}{2014}\natexlab{}.
\newblock \showarticletitle{Opentuner: An extensible framework for program
  autotuning}. In \bibinfo{booktitle}{\emph{PACT}}. \bibinfo{pages}{303--316}.
\newblock


\bibitem[\protect\citeauthoryear{Bradski}{Bradski}{2000}]%
        {opencv_library}
\bibfield{author}{\bibinfo{person}{Gary Bradski}.}
  \bibinfo{year}{2000}\natexlab{}.
\newblock \showarticletitle{The opencv library}.
\newblock \bibinfo{journal}{\emph{Dr Dobb's J. Software Tools}}
  \bibinfo{volume}{25} (\bibinfo{year}{2000}), \bibinfo{pages}{120--125}.
\newblock


\bibitem[\protect\citeauthoryear{{Cadence Stratus High-Level
  Synthesis}}{{Cadence Stratus High-Level Synthesis}}{[n.d.]}]%
        {cadence}
\bibfield{author}{\bibinfo{person}{{Cadence Stratus High-Level Synthesis}}.}
  \bibinfo{year}{[n.d.]}\natexlab{}.
\newblock
  \bibinfo{title}{\url{https://www.cadence.com/en_US/home/tools/digital-design-and-signoff/synthesis/stratus-high-level-synthesis.html}}.
\newblock
\newblock


\bibitem[\protect\citeauthoryear{{Catapult High-Level Synthesis}}{{Catapult
  High-Level Synthesis}}{[n.d.]}]%
        {catapult-hls}
\bibfield{author}{\bibinfo{person}{{Catapult High-Level Synthesis}}.}
  \bibinfo{year}{[n.d.]}\natexlab{}.
\newblock
  \bibinfo{title}{\url{https://eda.sw.siemens.com/en-US/ic/ic-design/high-level-synthesis-and-verification-platform/}}.
\newblock
\newblock


\bibitem[\protect\citeauthoryear{Che, Boyer, Meng, Tarjan, Sheaffer, Lee, and
  Skadron}{Che et~al\mbox{.}}{2009}]%
        {che2009rodinia}
\bibfield{author}{\bibinfo{person}{Shuai Che}, \bibinfo{person}{Michael Boyer},
  \bibinfo{person}{Jiayuan Meng}, \bibinfo{person}{David Tarjan},
  \bibinfo{person}{Jeremy~W Sheaffer}, \bibinfo{person}{Sang-Ha Lee}, {and}
  \bibinfo{person}{Kevin Skadron}.} \bibinfo{year}{2009}\natexlab{}.
\newblock \showarticletitle{Rodinia: A benchmark suite for heterogeneous
  computing}. In \bibinfo{booktitle}{\emph{IISWC}}. \bibinfo{pages}{44--54}.
\newblock


\bibitem[\protect\citeauthoryear{Chi, Cong, Wei, and Zhou}{Chi
  et~al\mbox{.}}{2018}]%
        {chi2018soda}
\bibfield{author}{\bibinfo{person}{Yuze Chi}, \bibinfo{person}{Jason Cong},
  \bibinfo{person}{Peng Wei}, {and} \bibinfo{person}{Peipei Zhou}.}
  \bibinfo{year}{2018}\natexlab{}.
\newblock \showarticletitle{SODA: stencil with optimized dataflow
  architecture}. In \bibinfo{booktitle}{\emph{ICCAD}}. \bibinfo{pages}{1--8}.
\newblock


\bibitem[\protect\citeauthoryear{Choi and Cong}{Choi and Cong}{2018}]%
        {choi2018hls}
\bibfield{author}{\bibinfo{person}{Young-kyu Choi} {and} \bibinfo{person}{Jason
  Cong}.} \bibinfo{year}{2018}\natexlab{}.
\newblock \showarticletitle{HLS-based optimization and design space exploration
  for applications with variable loop bounds}. In
  \bibinfo{booktitle}{\emph{ICCAD}}. \bibinfo{pages}{1--8}.
\newblock


\bibitem[\protect\citeauthoryear{Cong, Huang, Pan, Wang, and Zhang}{Cong
  et~al\mbox{.}}{2016a}]%
        {merlin}
\bibfield{author}{\bibinfo{person}{Jason Cong}, \bibinfo{person}{Muhuan Huang},
  \bibinfo{person}{Peichen Pan}, \bibinfo{person}{Yuxin Wang}, {and}
  \bibinfo{person}{Peng Zhang}.} \bibinfo{year}{2016}\natexlab{a}.
\newblock \showarticletitle{Source-to-source optimization for HLS}.
\newblock In \bibinfo{booktitle}{\emph{FPGAs for Software Programmers}}.
  \bibinfo{pages}{137--163}.
\newblock


\bibitem[\protect\citeauthoryear{Cong, Huang, Pan, Wu, and Zhang}{Cong
  et~al\mbox{.}}{2016b}]%
        {merlin_islped}
\bibfield{author}{\bibinfo{person}{Jason Cong}, \bibinfo{person}{Muhuan Huang},
  \bibinfo{person}{Peichen Pan}, \bibinfo{person}{Di Wu}, {and}
  \bibinfo{person}{Peng Zhang}.} \bibinfo{year}{2016}\natexlab{b}.
\newblock \showarticletitle{Software infrastructure for enabling FPGA-based
  accelerations in data centers}. In \bibinfo{booktitle}{\emph{ISLPED}}.
  \bibinfo{pages}{154--155}.
\newblock


\bibitem[\protect\citeauthoryear{Cong, Liu, Neuendorffer, Noguera, Vissers, and
  Zhang}{Cong et~al\mbox{.}}{2011}]%
        {cong11}
\bibfield{author}{\bibinfo{person}{Jason Cong}, \bibinfo{person}{Bin Liu},
  \bibinfo{person}{Stephen Neuendorffer}, \bibinfo{person}{Juanjo Noguera},
  \bibinfo{person}{Kees Vissers}, {and} \bibinfo{person}{Zhiru Zhang}.}
  \bibinfo{year}{2011}\natexlab{}.
\newblock \showarticletitle{High-level synthesis for FPGAs: From prototyping to
  deployment}. In \bibinfo{booktitle}{\emph{TCAD}}, Vol.~\bibinfo{volume}{30}.
  \bibinfo{pages}{473--491}.
\newblock


\bibitem[\protect\citeauthoryear{Cong and Wang}{Cong and Wang}{2018}]%
        {cong2018polysa}
\bibfield{author}{\bibinfo{person}{Jason Cong} {and} \bibinfo{person}{Jie
  Wang}.} \bibinfo{year}{2018}\natexlab{}.
\newblock \showarticletitle{PolySA: polyhedral-based systolic array
  auto-compilation}. In \bibinfo{booktitle}{\emph{ICCAD}}.
  \bibinfo{pages}{1--8}.
\newblock


\bibitem[\protect\citeauthoryear{Cong, Wei, Yu, and Zhang}{Cong
  et~al\mbox{.}}{2018}]%
        {autoaccel}
\bibfield{author}{\bibinfo{person}{Jason Cong}, \bibinfo{person}{Peng Wei},
  \bibinfo{person}{Cody~Hao Yu}, {and} \bibinfo{person}{Peng Zhang}.}
  \bibinfo{year}{2018}\natexlab{}.
\newblock \showarticletitle{Automated accelerator generation and optimization
  with composable, parallel and pipeline architecture}. In
  \bibinfo{booktitle}{\emph{DAC}}.
\newblock


\bibitem[\protect\citeauthoryear{{CyberWorkBench}}{{CyberWorkBench}}{[n.d.]}]%
        {nec}
\bibfield{author}{\bibinfo{person}{{CyberWorkBench}}.}
  \bibinfo{year}{[n.d.]}\natexlab{}.
\newblock
  \bibinfo{title}{\url{https://www.nec.com/en/global/prod/cwb/index.html}}.
\newblock
\newblock


\bibitem[\protect\citeauthoryear{Dagum and Menon}{Dagum and Menon}{1998}]%
        {dagum1998openmp}
\bibfield{author}{\bibinfo{person}{Leonardo Dagum} {and}
  \bibinfo{person}{Ramesh Menon}.} \bibinfo{year}{1998}\natexlab{}.
\newblock \showarticletitle{OpenMP: an industry standard API for shared-memory
  programming}.
\newblock \bibinfo{journal}{\emph{IEEE computational science and engineering}}
  \bibinfo{volume}{5}, \bibinfo{number}{1} (\bibinfo{year}{1998}),
  \bibinfo{pages}{46--55}.
\newblock


\bibitem[\protect\citeauthoryear{Dai, Zhou, Zhang, Ustun, Young, and Zhang}{Dai
  et~al\mbox{.}}{2018}]%
        {dai2018fast}
\bibfield{author}{\bibinfo{person}{Steve Dai}, \bibinfo{person}{Yuan Zhou},
  \bibinfo{person}{Hang Zhang}, \bibinfo{person}{Ecenur Ustun},
  \bibinfo{person}{Evangeline~FY Young}, {and} \bibinfo{person}{Zhiru Zhang}.}
  \bibinfo{year}{2018}\natexlab{}.
\newblock \showarticletitle{Fast and accurate estimation of quality of results
  in high-level synthesis with machine learning}. In
  \bibinfo{booktitle}{\emph{2018 IEEE 26th Annual International Symposium on
  Field-Programmable Custom Computing Machines (FCCM)}}. IEEE,
  \bibinfo{pages}{129--132}.
\newblock


\bibitem[\protect\citeauthoryear{Dennard, Gaensslen, Yu, Rideout, Bassous, and
  LeBlanc}{Dennard et~al\mbox{.}}{1974}]%
        {dennard74}
\bibfield{author}{\bibinfo{person}{Robert~H Dennard}, \bibinfo{person}{Fritz~H
  Gaensslen}, \bibinfo{person}{Hwa-Nien Yu}, \bibinfo{person}{V~Leo Rideout},
  \bibinfo{person}{Ernest Bassous}, {and} \bibinfo{person}{Andre~R LeBlanc}.}
  \bibinfo{year}{1974}\natexlab{}.
\newblock \showarticletitle{Design of ion-implanted MOSFET's with very small
  physical dimensions}.
\newblock \bibinfo{journal}{\emph{IEEE Journal of Solid-State Circuits}}
  \bibinfo{volume}{9}, \bibinfo{number}{5}, \bibinfo{pages}{256--268}.
\newblock


\bibitem[\protect\citeauthoryear{Duarte, Han, Harris, Jindariani, Kreinar,
  Kreis, Ngadiuba, Pierini, Rivera, Tran, et~al\mbox{.}}{Duarte
  et~al\mbox{.}}{2018}]%
        {hls4ml}
\bibfield{author}{\bibinfo{person}{Javier Duarte}, \bibinfo{person}{Song Han},
  \bibinfo{person}{Philip Harris}, \bibinfo{person}{Sergo Jindariani},
  \bibinfo{person}{Edward Kreinar}, \bibinfo{person}{Benjamin Kreis},
  \bibinfo{person}{Jennifer Ngadiuba}, \bibinfo{person}{Maurizio Pierini},
  \bibinfo{person}{Ryan Rivera}, \bibinfo{person}{Nhan Tran}, {et~al\mbox{.}}}
  \bibinfo{year}{2018}\natexlab{}.
\newblock \showarticletitle{Fast inference of deep neural networks in FPGAs for
  particle physics}.
\newblock \bibinfo{journal}{\emph{Journal of Instrumentation}}
  \bibinfo{volume}{13}, \bibinfo{number}{07} (\bibinfo{year}{2018}),
  \bibinfo{pages}{P07027}.
\newblock


\bibitem[\protect\citeauthoryear{{Falcon Computing Solutions, Inc}}{{Falcon
  Computing Solutions, Inc}}{[n.d.]}]%
        {fcs}
\bibfield{author}{\bibinfo{person}{{Falcon Computing Solutions, Inc}}.}
  \bibinfo{year}{[n.d.]}\natexlab{}.
\newblock \bibinfo{title}{\url{http://www.falcon-computing.com}}.
\newblock
\newblock


\bibitem[\protect\citeauthoryear{Ferretti, Ansaloni, and Pozzi}{Ferretti
  et~al\mbox{.}}{2018a}]%
        {ferretti2018cluster}
\bibfield{author}{\bibinfo{person}{Lorenzo Ferretti}, \bibinfo{person}{Giovanni
  Ansaloni}, {and} \bibinfo{person}{Laura Pozzi}.}
  \bibinfo{year}{2018}\natexlab{a}.
\newblock \showarticletitle{Cluster-based heuristic for high level synthesis
  design space exploration}.
\newblock \bibinfo{journal}{\emph{IEEE Transactions on Emerging Topics in
  Computing}}.
\newblock


\bibitem[\protect\citeauthoryear{Ferretti, Ansaloni, and Pozzi}{Ferretti
  et~al\mbox{.}}{2018b}]%
        {ferretti2018lattice}
\bibfield{author}{\bibinfo{person}{Lorenzo Ferretti}, \bibinfo{person}{Giovanni
  Ansaloni}, {and} \bibinfo{person}{Laura Pozzi}.}
  \bibinfo{year}{2018}\natexlab{b}.
\newblock \showarticletitle{Lattice-traversing design space exploration for
  high level synthesis}. In \bibinfo{booktitle}{\emph{ICCD}}.
  \bibinfo{pages}{210--217}.
\newblock


\bibitem[\protect\citeauthoryear{Fialho, Da~Costa, Schoenauer, and
  Sebag}{Fialho et~al\mbox{.}}{2010}]%
        {Fialho2010}
\bibfield{author}{\bibinfo{person}{{\'A}lvaro Fialho}, \bibinfo{person}{Luis
  Da~Costa}, \bibinfo{person}{Marc Schoenauer}, {and}
  \bibinfo{person}{Mich{\`e}le Sebag}.} \bibinfo{year}{2010}\natexlab{}.
\newblock \showarticletitle{Analyzing bandit-based adaptive operator selection
  mechanisms}.
\newblock \bibinfo{journal}{\emph{Annals of Mathematics and Artificial
  Intelligence}} \bibinfo{volume}{60}, \bibinfo{number}{1-2},
  \bibinfo{pages}{25--64}.
\newblock


\bibitem[\protect\citeauthoryear{{Intel}}{{Intel}}{[n.d.]}]%
        {vtune}
\bibfield{author}{\bibinfo{person}{{Intel}}.}
  \bibinfo{year}{[n.d.]}\natexlab{}.
\newblock
  \bibinfo{title}{\url{https://software.intel.com/content/www/us/en/develop/tools/oneapi/components/vtune-profiler.html}}.
\newblock
\newblock


\bibitem[\protect\citeauthoryear{{Intel SDK for OpenCL Applications}}{{Intel
  SDK for OpenCL Applications}}{[n.d.]}]%
        {intel-sdk}
\bibfield{author}{\bibinfo{person}{{Intel SDK for OpenCL Applications}}.}
  \bibinfo{year}{[n.d.]}\natexlab{}.
\newblock \bibinfo{title}{\url{https://software.intel.com/en-us/intel-opencl}}.
\newblock
\newblock


\bibitem[\protect\citeauthoryear{Koeplinger, Prabhakar, Zhang, Delimitrou,
  Kozyrakis, and Olukotun}{Koeplinger et~al\mbox{.}}{2016}]%
        {koeplinger2016automatic}
\bibfield{author}{\bibinfo{person}{David Koeplinger}, \bibinfo{person}{Raghu
  Prabhakar}, \bibinfo{person}{Yaqi Zhang}, \bibinfo{person}{Christina
  Delimitrou}, \bibinfo{person}{Christos Kozyrakis}, {and}
  \bibinfo{person}{Kunle Olukotun}.} \bibinfo{year}{2016}\natexlab{}.
\newblock \showarticletitle{Automatic generation of efficient accelerators for
  reconfigurable hardware}. In \bibinfo{booktitle}{\emph{ISCA}}.
  \bibinfo{pages}{115--127}.
\newblock


\bibitem[\protect\citeauthoryear{Krizhevsky, Sutskever, and Hinton}{Krizhevsky
  et~al\mbox{.}}{2012}]%
        {alexnet}
\bibfield{author}{\bibinfo{person}{Alex Krizhevsky}, \bibinfo{person}{Ilya
  Sutskever}, {and} \bibinfo{person}{Geoffrey~E Hinton}.}
  \bibinfo{year}{2012}\natexlab{}.
\newblock \showarticletitle{Imagenet classification with deep convolutional
  neural networks}. In \bibinfo{booktitle}{\emph{NIPS}}.
  \bibinfo{pages}{1097--1105}.
\newblock


\bibitem[\protect\citeauthoryear{Lai, Chi, Hu, Wang, Yu, Zhou, Cong, and
  Zhang}{Lai et~al\mbox{.}}{2019}]%
        {lai2019heterocl}
\bibfield{author}{\bibinfo{person}{Yi-Hsiang Lai}, \bibinfo{person}{Yuze Chi},
  \bibinfo{person}{Yuwei Hu}, \bibinfo{person}{Jie Wang},
  \bibinfo{person}{Cody~Hao Yu}, \bibinfo{person}{Yuan Zhou},
  \bibinfo{person}{Jason Cong}, {and} \bibinfo{person}{Zhiru Zhang}.}
  \bibinfo{year}{2019}\natexlab{}.
\newblock \showarticletitle{HeteroCL: A multi-paradigm programming
  infrastructure for software-defined reconfigurable computing}. In
  \bibinfo{booktitle}{\emph{Proceedings of the 2019 ACM/SIGDA International
  Symposium on Field-Programmable Gate Arrays}}. \bibinfo{pages}{242--251}.
\newblock


\bibitem[\protect\citeauthoryear{Liu and Carloni}{Liu and Carloni}{2013}]%
        {liu2013learning}
\bibfield{author}{\bibinfo{person}{Hung-Yi Liu} {and} \bibinfo{person}{Luca~P
  Carloni}.} \bibinfo{year}{2013}\natexlab{}.
\newblock \showarticletitle{On learning-based methods for design-space
  exploration with high-level synthesis}. In \bibinfo{booktitle}{\emph{DAC}}.
  \bibinfo{pages}{1--7}.
\newblock


\bibitem[\protect\citeauthoryear{Liu, Lau, and Schafer}{Liu
  et~al\mbox{.}}{2019}]%
        {liu2019accelerating}
\bibfield{author}{\bibinfo{person}{Shuangnan Liu}, \bibinfo{person}{Francis~CM
  Lau}, {and} \bibinfo{person}{Benjamin~Carrion Schafer}.}
  \bibinfo{year}{2019}\natexlab{}.
\newblock \showarticletitle{Accelerating fpga prototyping through predictive
  model-based hls design space exploration}. In
  \bibinfo{booktitle}{\emph{DAC}}. \bibinfo{pages}{1--6}.
\newblock


\bibitem[\protect\citeauthoryear{Mahapatra and Schafer}{Mahapatra and
  Schafer}{2014}]%
        {mahapatra2014machine}
\bibfield{author}{\bibinfo{person}{Anushree Mahapatra} {and}
  \bibinfo{person}{Benjamin~Carrion Schafer}.} \bibinfo{year}{2014}\natexlab{}.
\newblock \showarticletitle{Machine-learning based simulated annealer method
  for high level synthesis design space exploration}. In
  \bibinfo{booktitle}{\emph{ESLsyn}}. \bibinfo{pages}{1--6}.
\newblock


\bibitem[\protect\citeauthoryear{Nigam, Atapattu, Thomas, Li, Bauer, Ye, Koti,
  Sampson, and Zhang}{Nigam et~al\mbox{.}}{2020}]%
        {nigam2020predictable}
\bibfield{author}{\bibinfo{person}{Rachit Nigam}, \bibinfo{person}{Sachille
  Atapattu}, \bibinfo{person}{Samuel Thomas}, \bibinfo{person}{Zhijing Li},
  \bibinfo{person}{Theodore Bauer}, \bibinfo{person}{Yuwei Ye},
  \bibinfo{person}{Apurva Koti}, \bibinfo{person}{Adrian Sampson}, {and}
  \bibinfo{person}{Zhiru Zhang}.} \bibinfo{year}{2020}\natexlab{}.
\newblock \showarticletitle{Predictable accelerator design with time-sensitive
  affine types}.
\newblock \bibinfo{journal}{\emph{arXiv preprint arXiv:2004.04852}}
  (\bibinfo{year}{2020}).
\newblock


\bibitem[\protect\citeauthoryear{Prabhakar, Koeplinger, Brown, Lee, De~Sa,
  Kozyrakis, and Olukotun}{Prabhakar et~al\mbox{.}}{2016}]%
        {prabhakar2016generating}
\bibfield{author}{\bibinfo{person}{Raghu Prabhakar}, \bibinfo{person}{David
  Koeplinger}, \bibinfo{person}{Kevin~J Brown}, \bibinfo{person}{HyoukJoong
  Lee}, \bibinfo{person}{Christopher De~Sa}, \bibinfo{person}{Christos
  Kozyrakis}, {and} \bibinfo{person}{Kunle Olukotun}.}
  \bibinfo{year}{2016}\natexlab{}.
\newblock \showarticletitle{Generating configurable hardware from parallel
  patterns}.
\newblock \bibinfo{journal}{\emph{ASPLOS}} \bibinfo{volume}{51},
  \bibinfo{number}{4}, \bibinfo{pages}{651--665}.
\newblock


\bibitem[\protect\citeauthoryear{Putnam, Caulfield, Chung, Chiou,
  Constantinides, Demme, Esmaeilzadeh, Fowers, Gopal, Gray,
  et~al\mbox{.}}{Putnam et~al\mbox{.}}{2014}]%
        {catapult}
\bibfield{author}{\bibinfo{person}{Andrew Putnam}, \bibinfo{person}{Adrian~M
  Caulfield}, \bibinfo{person}{Eric~S Chung}, \bibinfo{person}{Derek Chiou},
  \bibinfo{person}{Kypros Constantinides}, \bibinfo{person}{John Demme},
  \bibinfo{person}{Hadi Esmaeilzadeh}, \bibinfo{person}{Jeremy Fowers},
  \bibinfo{person}{Gopi~Prashanth Gopal}, \bibinfo{person}{Jan Gray},
  {et~al\mbox{.}}} \bibinfo{year}{2014}\natexlab{}.
\newblock \showarticletitle{A reconfigurable fabric for accelerating
  large-scale datacenter services}. In \bibinfo{booktitle}{\emph{ISCA}}.
  \bibinfo{pages}{13--24}.
\newblock


\bibitem[\protect\citeauthoryear{Reagen, Adolf, Shao, Wei, and Brooks}{Reagen
  et~al\mbox{.}}{2014}]%
        {machsuite}
\bibfield{author}{\bibinfo{person}{Brandon Reagen}, \bibinfo{person}{Robert
  Adolf}, \bibinfo{person}{Yakun~Sophia Shao}, \bibinfo{person}{Gu-Yeon Wei},
  {and} \bibinfo{person}{David Brooks}.} \bibinfo{year}{2014}\natexlab{}.
\newblock \showarticletitle{Machsuite: Benchmarks for accelerator design and
  customized architectures}. In \bibinfo{booktitle}{\emph{IISWC}}.
  \bibinfo{pages}{110--119}.
\newblock


\bibitem[\protect\citeauthoryear{Reggiani, Rabozzi, Nestorov, Scolari,
  Stornaiuolo, and Santambrogio}{Reggiani et~al\mbox{.}}{2019}]%
        {reggiani2019pareto}
\bibfield{author}{\bibinfo{person}{Enrico Reggiani}, \bibinfo{person}{Marco
  Rabozzi}, \bibinfo{person}{Anna~Maria Nestorov}, \bibinfo{person}{Alberto
  Scolari}, \bibinfo{person}{Luca Stornaiuolo}, {and} \bibinfo{person}{Marco
  Santambrogio}.} \bibinfo{year}{2019}\natexlab{}.
\newblock \showarticletitle{Pareto optimal design space exploration for
  accelerated CNN on FPGA}. In \bibinfo{booktitle}{\emph{IPDPSW}}.
  \bibinfo{pages}{107--114}.
\newblock


\bibitem[\protect\citeauthoryear{Schafer}{Schafer}{2017}]%
        {schafer2017parallel}
\bibfield{author}{\bibinfo{person}{Benjamin~Carrion Schafer}.}
  \bibinfo{year}{2017}\natexlab{}.
\newblock \showarticletitle{Parallel high-level synthesis design space
  exploration for behavioral ips of exact latencies}.
\newblock \bibinfo{journal}{\emph{TODAES}} \bibinfo{volume}{22},
  \bibinfo{number}{4}, \bibinfo{pages}{1--20}.
\newblock


\bibitem[\protect\citeauthoryear{Schafer and Wakabayashi}{Schafer and
  Wakabayashi}{2012a}]%
        {schafer2012divide}
\bibfield{author}{\bibinfo{person}{Benjamin~Carrion Schafer} {and}
  \bibinfo{person}{Kazutoshi Wakabayashi}.} \bibinfo{year}{2012}\natexlab{a}.
\newblock \showarticletitle{Divide and conquer high-level synthesis design
  space exploration}.
\newblock \bibinfo{journal}{\emph{TODAES}} \bibinfo{volume}{17},
  \bibinfo{number}{3}, \bibinfo{pages}{1--19}.
\newblock


\bibitem[\protect\citeauthoryear{Schafer and Wakabayashi}{Schafer and
  Wakabayashi}{2012b}]%
        {schafer2012machine}
\bibfield{author}{\bibinfo{person}{B~Carrion Schafer} {and}
  \bibinfo{person}{Kazutoshi Wakabayashi}.} \bibinfo{year}{2012}\natexlab{b}.
\newblock \showarticletitle{Machine learning predictive modelling high-level
  synthesis design space exploration}. In \bibinfo{booktitle}{\emph{IET
  computers \& digital techniques}}, Vol.~\bibinfo{volume}{6}.
  \bibinfo{pages}{153--159}.
\newblock


\bibitem[\protect\citeauthoryear{Snoek, Rippel, Swersky, Kiros, Satish,
  Sundaram, Patwary, Prabhat, and Adams}{Snoek et~al\mbox{.}}{2015}]%
        {snoek2015scalable}
\bibfield{author}{\bibinfo{person}{Jasper Snoek}, \bibinfo{person}{Oren
  Rippel}, \bibinfo{person}{Kevin Swersky}, \bibinfo{person}{Ryan Kiros},
  \bibinfo{person}{Nadathur Satish}, \bibinfo{person}{Narayanan Sundaram},
  \bibinfo{person}{Mostofa Patwary}, \bibinfo{person}{Mr Prabhat}, {and}
  \bibinfo{person}{Ryan Adams}.} \bibinfo{year}{2015}\natexlab{}.
\newblock \showarticletitle{Scalable bayesian optimization using deep neural
  networks}. In \bibinfo{booktitle}{\emph{International conference on machine
  learning}}. PMLR, \bibinfo{pages}{2171--2180}.
\newblock


\bibitem[\protect\citeauthoryear{Sohrabizadeh, Wang, and Cong}{Sohrabizadeh
  et~al\mbox{.}}{2020}]%
        {sohrabizadeh2020end}
\bibfield{author}{\bibinfo{person}{Atefeh Sohrabizadeh}, \bibinfo{person}{Jie
  Wang}, {and} \bibinfo{person}{Jason Cong}.} \bibinfo{year}{2020}\natexlab{}.
\newblock \showarticletitle{End-to-End Optimization of Deep Learning
  Applications}. In \bibinfo{booktitle}{\emph{FPGA}}.
  \bibinfo{pages}{133--139}.
\newblock


\bibitem[\protect\citeauthoryear{Sun, Chen, Liu, Miao, Chen, Yu, and Yu}{Sun
  et~al\mbox{.}}{2021}]%
        {date21}
\bibfield{author}{\bibinfo{person}{Qi Sun}, \bibinfo{person}{Tinghuan Chen},
  \bibinfo{person}{Siting Liu}, \bibinfo{person}{Jin Miao},
  \bibinfo{person}{Jianli Chen}, \bibinfo{person}{Hao Yu}, {and}
  \bibinfo{person}{Bei Yu}.} \bibinfo{year}{2021}\natexlab{}.
\newblock \showarticletitle{Correlated Multi-objective Multi-fidelity
  Optimization for HLS Directives Design}. In
  \bibinfo{booktitle}{\emph{IEEE/ACM Proceedings Design, Automation and Test in
  Europe (DATE)}}. \bibinfo{pages}{01--05}.
\newblock


\bibitem[\protect\citeauthoryear{Vandebon, Coutinho, Luk, Nurvitadhi, and
  Todman}{Vandebon et~al\mbox{.}}{2020}]%
        {vandebon2020artisan}
\bibfield{author}{\bibinfo{person}{Jessica Vandebon}, \bibinfo{person}{Jose~GF
  Coutinho}, \bibinfo{person}{Wayne Luk}, \bibinfo{person}{Eriko Nurvitadhi},
  {and} \bibinfo{person}{Tim Todman}.} \bibinfo{year}{2020}\natexlab{}.
\newblock \showarticletitle{Artisan: a Meta-Programming Approach For Codifying
  Optimisation Strategies}. In \bibinfo{booktitle}{\emph{FCCM}}.
  \bibinfo{pages}{177--185}.
\newblock


\bibitem[\protect\citeauthoryear{{Vivado HLS}}{{Vivado HLS}}{[n.d.]}]%
        {hls}
\bibfield{author}{\bibinfo{person}{{Vivado HLS}}.}
  \bibinfo{year}{[n.d.]}\natexlab{}.
\newblock \bibinfo{title}{\url{www.xilinx.com/products/design-tools/vivado}}.
\newblock
\newblock


\bibitem[\protect\citeauthoryear{Wang, Guo, and Cong}{Wang
  et~al\mbox{.}}{2021}]%
        {autosa}
\bibfield{author}{\bibinfo{person}{Jie Wang}, \bibinfo{person}{Licheng Guo},
  {and} \bibinfo{person}{Jason Cong}.} \bibinfo{year}{2021}\natexlab{}.
\newblock \showarticletitle{AutoSA: A Polyhedral Compiler for High-Performance
  Systolic Arrays on FPGA}. In \bibinfo{booktitle}{\emph{Proceedings of the
  2021 ACM/SIGDA international symposium on Field-programmable gate arrays}}.
\newblock


\bibitem[\protect\citeauthoryear{Wang, Liang, and Zhang}{Wang
  et~al\mbox{.}}{2017}]%
        {wang2017flexcl}
\bibfield{author}{\bibinfo{person}{Shuo Wang}, \bibinfo{person}{Yun Liang},
  {and} \bibinfo{person}{Wei Zhang}.} \bibinfo{year}{2017}\natexlab{}.
\newblock \showarticletitle{Flexcl: An analytical performance model for opencl
  workloads on flexible fpgas}. In \bibinfo{booktitle}{\emph{DAC}}.
  \bibinfo{pages}{1--6}.
\newblock


\bibitem[\protect\citeauthoryear{{Xilinx}}{{Xilinx}}{[n.d.]}]%
        {acquisition}
\bibfield{author}{\bibinfo{person}{{Xilinx}}.}
  \bibinfo{year}{[n.d.]}\natexlab{}.
\newblock
  \bibinfo{title}{\url{https://www.xilinx.com/about/xilinx-ventures/falcon-computing.html}}.
\newblock
\newblock


\bibitem[\protect\citeauthoryear{{Xilinx Vitis Libraries}}{{Xilinx Vitis
  Libraries}}{[n.d.]}]%
        {vitis}
\bibfield{author}{\bibinfo{person}{{Xilinx Vitis Libraries}}.}
  \bibinfo{year}{[n.d.]}\natexlab{}.
\newblock \bibinfo{title}{\url{www.github.com/Xilinx/Vitis\_Libraries}}.
\newblock
\newblock


\bibitem[\protect\citeauthoryear{{Xilinx Vitis Platform}}{{Xilinx Vitis
  Platform}}{[n.d.]}]%
        {vitis-platform}
\bibfield{author}{\bibinfo{person}{{Xilinx Vitis Platform}}.}
  \bibinfo{year}{[n.d.]}\natexlab{}.
\newblock
  \bibinfo{title}{\url{https://www.xilinx.com/products/design-tools/vitis/vitis-platform.html}}.
\newblock
\newblock


\bibitem[\protect\citeauthoryear{Xu, Liu, Zhao, Yang, Luo, and Zhang}{Xu
  et~al\mbox{.}}{2017}]%
        {datuner}
\bibfield{author}{\bibinfo{person}{Chang Xu}, \bibinfo{person}{Gai Liu},
  \bibinfo{person}{Ritchie Zhao}, \bibinfo{person}{Stephen Yang},
  \bibinfo{person}{Guojie Luo}, {and} \bibinfo{person}{Zhiru Zhang}.}
  \bibinfo{year}{2017}\natexlab{}.
\newblock \showarticletitle{A parallel bandit-based approach for autotuning
  fpga compilation}. In \bibinfo{booktitle}{\emph{FPGA}}.
  \bibinfo{pages}{157--166}.
\newblock


\bibitem[\protect\citeauthoryear{Xu, Zhang, Hao, Zhao, Zhang, Wang, Li, Guan,
  Chen, and Lin}{Xu et~al\mbox{.}}{2020}]%
        {xu2020autodnnchip}
\bibfield{author}{\bibinfo{person}{Pengfei Xu}, \bibinfo{person}{Xiaofan
  Zhang}, \bibinfo{person}{Cong Hao}, \bibinfo{person}{Yang Zhao},
  \bibinfo{person}{Yongan Zhang}, \bibinfo{person}{Yue Wang},
  \bibinfo{person}{Chaojian Li}, \bibinfo{person}{Zetong Guan},
  \bibinfo{person}{Deming Chen}, {and} \bibinfo{person}{Yingyan Lin}.}
  \bibinfo{year}{2020}\natexlab{}.
\newblock \showarticletitle{AutoDNNchip: An automated dnn chip predictor and
  builder for both FPGAs and ASICs}. In \bibinfo{booktitle}{\emph{FPGA}}.
  \bibinfo{pages}{40--50}.
\newblock


\bibitem[\protect\citeauthoryear{Xydis, Palermo, Zaccaria, and Silvano}{Xydis
  et~al\mbox{.}}{2014}]%
        {xydis2015spirit}
\bibfield{author}{\bibinfo{person}{Sotirios Xydis}, \bibinfo{person}{Gianluca
  Palermo}, \bibinfo{person}{Vittorio Zaccaria}, {and}
  \bibinfo{person}{Cristina Silvano}.} \bibinfo{year}{2014}\natexlab{}.
\newblock \showarticletitle{SPIRIT: Spectral-Aware pareto iterative refinement
  optimization for supervised high-level synthesis}. In
  \bibinfo{booktitle}{\emph{TCAD}}, Vol.~\bibinfo{volume}{34}.
  \bibinfo{pages}{155--159}.
\newblock


\bibitem[\protect\citeauthoryear{Yu, Wei, Grossman, Zhang, Sarker, and Cong}{Yu
  et~al\mbox{.}}{2018}]%
        {s2fa}
\bibfield{author}{\bibinfo{person}{Cody~Hao Yu}, \bibinfo{person}{Peng Wei},
  \bibinfo{person}{Max Grossman}, \bibinfo{person}{Peng Zhang},
  \bibinfo{person}{Vivek Sarker}, {and} \bibinfo{person}{Jason Cong}.}
  \bibinfo{year}{2018}\natexlab{}.
\newblock \showarticletitle{S2FA: an accelerator automation framework for
  heterogeneous computing in datacenters}. In \bibinfo{booktitle}{\emph{DAC}}.
  \bibinfo{pages}{1--6}.
\newblock


\bibitem[\protect\citeauthoryear{Zacharopoulos, Ferretti, Ansaloni,
  Di~Guglielmo, Carloni, and Pozzi}{Zacharopoulos et~al\mbox{.}}{2019}]%
        {zacharopoulos2019compiler}
\bibfield{author}{\bibinfo{person}{Georgios Zacharopoulos},
  \bibinfo{person}{Lorenzo Ferretti}, \bibinfo{person}{Giovanni Ansaloni},
  \bibinfo{person}{Giuseppe Di~Guglielmo}, \bibinfo{person}{Luca Carloni},
  {and} \bibinfo{person}{Laura Pozzi}.} \bibinfo{year}{2019}\natexlab{}.
\newblock \showarticletitle{Compiler-assisted selection of hardware
  acceleration candidates from application source code}. In
  \bibinfo{booktitle}{\emph{ICCD}}. \bibinfo{pages}{129--137}.
\newblock


\bibitem[\protect\citeauthoryear{Zhang, Fan, Jiang, Han, Yang, and Cong}{Zhang
  et~al\mbox{.}}{2008}]%
        {zhang2008autopilot}
\bibfield{author}{\bibinfo{person}{Zhiru Zhang}, \bibinfo{person}{Yiping Fan},
  \bibinfo{person}{Wei Jiang}, \bibinfo{person}{Guoling Han},
  \bibinfo{person}{Changqi Yang}, {and} \bibinfo{person}{Jason Cong}.}
  \bibinfo{year}{2008}\natexlab{}.
\newblock \showarticletitle{AutoPilot: A platform-based ESL synthesis system}.
\newblock In \bibinfo{booktitle}{\emph{High-Level Synthesis}}.
  \bibinfo{pages}{99--112}.
\newblock


\bibitem[\protect\citeauthoryear{Zhao, Feng, Sinha, Zhang, Liang, and He}{Zhao
  et~al\mbox{.}}{2017}]%
        {comba}
\bibfield{author}{\bibinfo{person}{Jieru Zhao}, \bibinfo{person}{Liang Feng},
  \bibinfo{person}{Sharad Sinha}, \bibinfo{person}{Wei Zhang},
  \bibinfo{person}{Yun Liang}, {and} \bibinfo{person}{Bingsheng He}.}
  \bibinfo{year}{2017}\natexlab{}.
\newblock \showarticletitle{COMBA: A comprehensive model-based analysis
  framework for high level synthesis of real applications}. In
  \bibinfo{booktitle}{\emph{ICCAD}}. \bibinfo{pages}{430--437}.
\newblock


\bibitem[\protect\citeauthoryear{Zheng, Liang, Wang, Chen, and Sheng}{Zheng
  et~al\mbox{.}}{2020}]%
        {zheng2020flextensor}
\bibfield{author}{\bibinfo{person}{Size Zheng}, \bibinfo{person}{Yun Liang},
  \bibinfo{person}{Shuo Wang}, \bibinfo{person}{Renze Chen}, {and}
  \bibinfo{person}{Kaiwen Sheng}.} \bibinfo{year}{2020}\natexlab{}.
\newblock \showarticletitle{FlexTensor: An Automatic Schedule Exploration and
  Optimization Framework for Tensor Computation on Heterogeneous System}. In
  \bibinfo{booktitle}{\emph{ASPLOS}}. \bibinfo{pages}{859--873}.
\newblock


\bibitem[\protect\citeauthoryear{Zhong, Prakash, Liang, Mitra, and Niar}{Zhong
  et~al\mbox{.}}{2016}]%
        {linanalyzer}
\bibfield{author}{\bibinfo{person}{Guanwen Zhong}, \bibinfo{person}{Alok
  Prakash}, \bibinfo{person}{Yun Liang}, \bibinfo{person}{Tulika Mitra}, {and}
  \bibinfo{person}{Smail Niar}.} \bibinfo{year}{2016}\natexlab{}.
\newblock \showarticletitle{Lin-analyzer: a high-level performance analysis
  tool for FPGA-based accelerators}. In \bibinfo{booktitle}{\emph{DAC}}.
  \bibinfo{pages}{1--6}.
\newblock


\bibitem[\protect\citeauthoryear{Zhong, Prakash, Wang, Liang, Mitra, and
  Niar}{Zhong et~al\mbox{.}}{2017}]%
        {zhong2017design}
\bibfield{author}{\bibinfo{person}{Guanwen Zhong}, \bibinfo{person}{Alok
  Prakash}, \bibinfo{person}{Siqi Wang}, \bibinfo{person}{Yun Liang},
  \bibinfo{person}{Tulika Mitra}, {and} \bibinfo{person}{Smail Niar}.}
  \bibinfo{year}{2017}\natexlab{}.
\newblock \showarticletitle{Design Space exploration of FPGA-based accelerators
  with multi-level parallelism}. In \bibinfo{booktitle}{\emph{DATE}}.
  \bibinfo{pages}{1141--1146}.
\newblock


\bibitem[\protect\citeauthoryear{Zhong, Venkataramani, Liang, Mitra, and
  Niar}{Zhong et~al\mbox{.}}{2014}]%
        {zhong2014design}
\bibfield{author}{\bibinfo{person}{Guanwen Zhong},
  \bibinfo{person}{Vanchinathan Venkataramani}, \bibinfo{person}{Yun Liang},
  \bibinfo{person}{Tulika Mitra}, {and} \bibinfo{person}{Smail Niar}.}
  \bibinfo{year}{2014}\natexlab{}.
\newblock \showarticletitle{Design space exploration of multiple loops on FPGAs
  using high level synthesis}. In \bibinfo{booktitle}{\emph{ICCD}}.
  \bibinfo{pages}{456--463}.
\newblock


\bibitem[\protect\citeauthoryear{Zohouri, Podobas, and Matsuoka}{Zohouri
  et~al\mbox{.}}{2018}]%
        {zohouri2018combined}
\bibfield{author}{\bibinfo{person}{Hamid~Reza Zohouri}, \bibinfo{person}{Artur
  Podobas}, {and} \bibinfo{person}{Satoshi Matsuoka}.}
  \bibinfo{year}{2018}\natexlab{}.
\newblock \showarticletitle{Combined spatial and temporal blocking for
  high-performance stencil computation on FPGAs using OpenCL}. In
  \bibinfo{booktitle}{\emph{Proceedings of the 2018 ACM/SIGDA International
  Symposium on Field-Programmable Gate Arrays}}. \bibinfo{pages}{153--162}.
\newblock


\bibitem[\protect\citeauthoryear{Zuo, Li, Chen, Pouchet, Zhong, and Cong}{Zuo
  et~al\mbox{.}}{2013}]%
        {polyframework}
\bibfield{author}{\bibinfo{person}{Wei Zuo}, \bibinfo{person}{Peng Li},
  \bibinfo{person}{Deming Chen}, \bibinfo{person}{Louis-No{\"e}l Pouchet},
  \bibinfo{person}{Shunan Zhong}, {and} \bibinfo{person}{Jason Cong}.}
  \bibinfo{year}{2013}\natexlab{}.
\newblock \showarticletitle{Improving polyhedral code generation for high-level
  synthesis}. In \bibinfo{booktitle}{\emph{CODES+ ISSS}}.
  \bibinfo{pages}{1--10}.
\newblock


\end{thebibliography}

\appendix

\section{Appendix}
\subsection{Optimized HLS Code for CNN} \label{a1_cnn}
Code~\ref{code:cnn_hls_tranformed_opt} shows the optimized HLS code for the CNN algorithm in Code~\ref{code:cnn_hls_tranformed} after applying the code transformations and pragmas listed in Table~\ref{tbl:cnn_analysis}.

\begin{lstlisting}[language=C,caption=Optimized CNN HLS C Code Snippet\label{code:cnn_hls_tranformed_opt},
     breaklines,breakatwhitespace]
// Skip const variable initizalization for brevity

void CnnKernel(const ap\_uint< 128 > * input, float weight, 
                const  ap\_uint< 512 > * bias, ap\_uint< 512 > * output){
#pragma HLS INTERFACE m_axi port=input bundle=gmem1 depth=3326977
#pragma HLS INTERFACE s_axilite port=input bundle=control
// Skip the rest for brevity

  float bias_buf[ParallelOut][ParallelOut];
#pragma HLS array_partition variable=bias_buf complete dim=2

  float C[ParallelOut][ImSize][ImSize];
#pragma HLS array_partition variable=C cyclic factor=8 dim=3
#pragma HLS array_partition variable=C cyclic factor=2 dim=2
#pragma HLS array_partition variable=C complete dim=1

  LoadBurst(bias, bias_buf);

  for (int i = 0; i < NumOut / ParallelOut; i++) {
    float weight_buf[NumOut / ParallelOut][NumIn][kKernel][kKernel];
#pragma HLS array_partition variable=weight_buf complete dim=4
#pragma HLS array_partition variable=weight_buf complete dim=3
#pragma HLS array_partition variable=weight_buf complete dim=1

    float output_buf[NumOut / ParallelOut][OutImSize][OutImSize];
#pragma HLS array_partition variable=output_buf cyclic factor=16 dim=3
#pragma HLS array_partition variable=output_buf complete dim=1

    LoadBurst(weight, weight_buf);
    // Initialization
    for (int h = 0; h < ImSize; ++h) {
      for (int w = 0; w < ImSize / 4; ++w) {
#pragma HLS dependence variable=C array inter false
#pragma HLS pipeline
        for (int w_sub = 0; w_sub < 4; ++w_sub) {
#pragma HLS unroll
          for (int po = 0; po < ParallelOut; po++) {           
#pragma HLS unroll
            C[po][h][w * 4 + w_sub] = 0.f;
    } } } } 
    // Convolution
    for (int j = 0; j < NumIn; ++j) {
      float input_buf[InImSize][InImSize];
#pragma HLS array_partition variable=input_buf cyclic factor=8 dim=2
#pragma HLS array_partition variable=input_buf cyclic factor=5 dim=1
      LoadBurst(input, input_buf);
      for (int h = 0; h < ImSize; ++h) {
        for (int w = 0; w < ImSize / 4; ++w) {
#pragma HLS dependence variable=C array inter false
#pragma HLS pipeline
          for (int w_sub = 0; w_sub < 4; ++w_sub) {
#pragma HLS unroll
            for (int po = 0; po < ParallelOut; po++) {
#pragma HLS unroll
              float tmp = 0.f;
              for (int p = 0; p < kKernel; ++p) {
#pragma HLS unroll
                for (int q = 0; q < kKernel; ++q) {
#pragma HLS unroll
                  tmp += ...;
              } }
              C[po][h][w * 4 + w_sub] += tmp;
    } } } } }
    // ReLU + Max pooling
    for (int h = 0; h < OutImSize; ++h) {
      for (int w = 0; w < OutImSize; ++w) {       
#pragma HLS dependence variable=output_buf array inter false
#pragma HLS pipeline
        for (int po = 0; po < ParallelOut; po++) {
#pragma HLS unroll
            output_buf(h, w, po) = ...
    } } }
    StoreBurst(output, output_buf);
} }
\end{lstlisting}
\subsection{Detailed Comparison to the Vitis Library} \label{sec:a2_vitis}
\begin{figure*}[!htb]
	\centering
	\includegraphics[width=\linewidth]{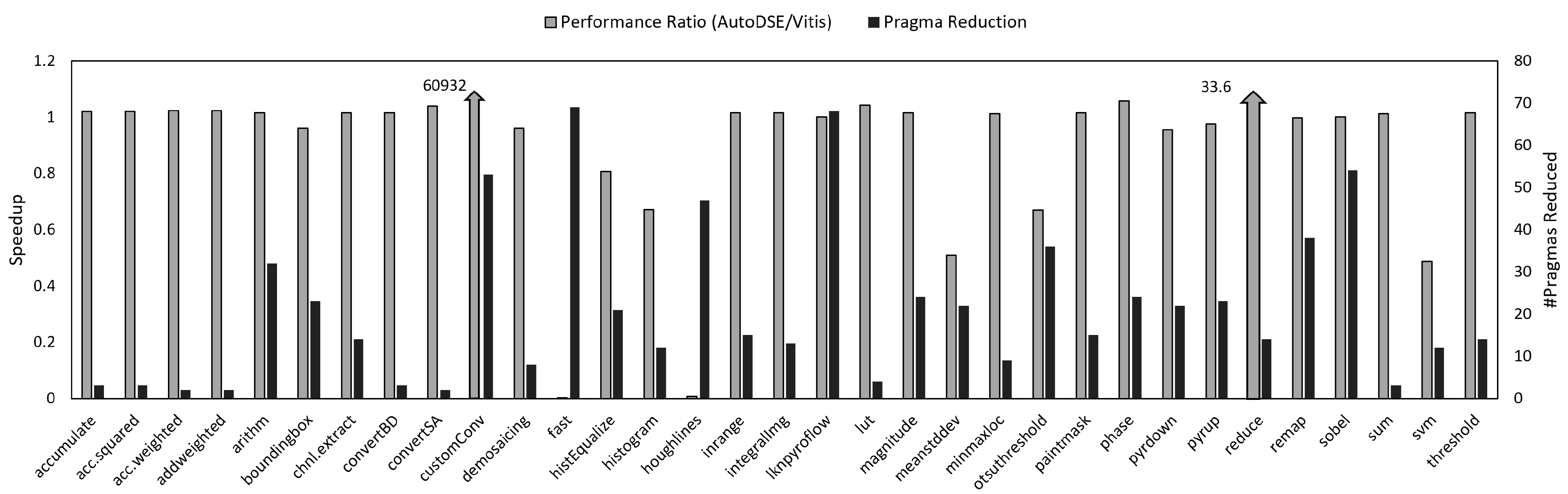} 
	\caption{Speedup and Number of Reduced Pragmas Using AutoDSE Compared to Vision Kernels of Xilinx Vitis libraries~\cite{vitis}}
	\label{fig:exp_vitis}
\end{figure*}

Fig.~\ref{fig:exp_vitis} depicts the detailed comparison of {$\framework$} to the expert-level manual HLS designs from Xilinx Vitis libraries~\cite{vitis}. As explained in Section~\ref{sec:exp_vitis}, when testing with {$\framework$}, all the optimization pragmas that the Merlin Compiler can derive with the help of its own pragmas are removed. {$\framework$} can achieve the same performance while using {\pragmareductionvitis} less pragmas, on the geometric mean.

\end{document}